\documentclass[11pt]{article}

\usepackage{amsmath}

\usepackage[cp1251]{inputenc}

\newcommand{\N}{N\raise.7ex\hbox{\underline{$\circ $}}$\;$}

\begin{document}

\begin{center}

{\bf  E.M. Ovsiyuk,   V.V.  Kisel,   V.M. Red'kov  \\[3mm]
ON A DIRAC PARTICLE IN AN UNIFORM MAGNETIC FIELD IN 3-DIMENSIONAL
SPACES OF CONSTANT CURVATURE
}\\[3mm]
Mosyr State Pedagogical University, Belarus\\
Institute of Physics, National Academy of Sciences  of Belarus\\
               e.ovsiyuk@mail.ru ; redkov@dragon.bas-net.by
\end{center}

\begin{quotation}

There are constructed exact solutions of the quantum-mecha\-nical
Dirac equation for a spin S=1/2 particle in Riemannian space of
constant negative curvature,
 hyperbolic Lobachevsky space,  in presence of an external magnetic field, analogue of
  the homogeneous magnetic field in the Minkowski  space. A generalized formula for energy levels,
   describing   quantization of the trans\-versal motion of the particle in magnetic field  has been obtained.
The same problem is solved for spin 1/2 particle
 in
the space of constant positive curvature, spherical Riemann space.
  A generalized formula for energy levels, describing quantization of the
  transversal and along the magnetic field motions of the particle on the back\-ground of the Riemann space
   geometry, is obtained.

   \end{quotation}

 Short title:       Spin 1/2 Particle in magnetic field, curved models

PACS: 02.40.Ky, 03.65Ge, 04.62.+v,

\section{ Introduction }

The  quantization  of a quantum-mechanical particle in the
homogeneous magnetic field belongs to classical  problems in
physics  \cite{1,2,3, Landau-Lifshitz}.
In 1985 -- 2010, a more general problem in a curved Riemannian background,  hyperbolic and
spherical  planes,  was extensively studied
\cite{Comtet-1985, Comtet-1987,
 Aoki-1987, Groshe-1988,
Klauder-Onofri,
Avron-Pnueli-1990,
Plyushchay-1991(1), Plyushchay-1991(2),
Dunne-1992,
Plyushchay-1995,
Alimohammadi-Shafei Deh Abad-1996, Alimohammadi-Mohseni Sadjadi-1996,
Onofri-2001,
Negro et al-2001,
Gamboa et al-2001,
Klishevich-Plyushchay-2002,
Drukker et  al-2004,
Ghanmi-Intissar-2004,
Correa-Jakubsky-Plyushchay-2009,
Alvarez-Cortes-Horvathy-Plyushchay-2009},
   providing us with a new system having intriguing
dynamics and symmetry, both on  classical and quantum levels.

 Extension  to 3-dimensional hyperbolic and spherical spaces was
 performed recently.
In  \cite{4,5,6}, exact solutions for  a scalar particle in
extended problem, particle in  external magnetic field on the
background of Lobachevsky $H_{3}$  and Riemann $S_{3}$ spatial
geometries were found. A corresponding system in the frames of
classical mechanics was examined in
 \cite{7,8,9}. In the present paper, we consider a quantum-mechanical problem  a particle with spin $1/2$ described
 by the Dirac equation in  3-dimensional Lobachevsky  and Riemann space models
 in presence of the external magnetic field.

The paper is organized as follows. In Section 2, the general
covariant Dirac equation is specified in Lobachevsky  model in
special cylindric coordinate system, in which a generalized
concept of  an homogeneous magnetic field can be defined
straightforwardly. Then the variables in the Dirac equation are
separated, and the problem is reduced to a couple of differential
equations, describing the motion of the Dirac particle in $z$ and
$r$-directions. In Section 3, differential equations in
$z$-variable are solved in terms of hypergeometric  functions, no
quantization arises in accordance with the  topological properties
of the Lobachevsky model. In Section 4, we examine differential
equations in radial variable $r$, producing  a generalized
quantization rule for transversal energy, due to the presence of
external magnetic field.

\section{  Cylindric coordinates and the Dirac equation in
hyperbolic space $H_{3}$, separation of the variables }

 In the Lobachevsky space, let us use an extended cylindric coordinates (see \cite{Olevsky})
\begin{eqnarray}
dS^{2} =  dt^{2} -  \mbox{ cosh}^{2} z ( d r^{2} + \mbox{sinh}^{2} r
\; d \phi^{2} ) - dz^{2}\; ,
\nonumber\\
u_{1} = \mbox{cosh} \; z \; \mbox{sinh}\; r \cos \phi \; , \;\; u_{2}
= \mbox{cosh} \; z \; \mbox{sinh}\;r \sin \phi \; ,
 \nonumber \\
 u_{3}
= \mbox{sinh}\; z \; , \;\;  u_{0} = \mbox{cosh}\; z \; \mbox{cosh} \;
r \; ; \label{A.1}
\end{eqnarray}

\noindent \noindent where  $x^{j}= (r,\; \phi,\; z )$: $ r \in [ 0
, + \infty )\; , \; \phi \in [ 0 , 2\pi ]\;, \; z \in (- \infty ,
+ \infty ) \;$; the curvature radius $\rho$ is taken as a  unit of
the length. An analogue of usual homogeneous magnetic field is
defined as \cite{4,5,6}
\begin{eqnarray}
 A_{\phi} = -2B \; \mbox{sinh}^{2} {r \over 2} = - B\; (\mbox{cosh}\; r -1 )\; .
\label{A.2}
\end{eqnarray}

\noindent To coordinates  (\ref{A.1})  there corresponds the
tetrad
\begin{eqnarray}
 e_{(a)}^{\beta} = \left |
\begin{array}{llll}
1 & 0 & 0 & 0 \\
0 & \mbox{cosh}^{-1}z & 0 & 0 \\
0 & 0 & \mbox{cosh}^{-1}z\;\mbox{sinh}^{-1} r & 0 \\
0 & 0 & 0 & 1
\end{array} \right |   .
\label{A.3}
\end{eqnarray}

\noindent  Christoffel symbols  $\Gamma^{r}_{\;\;jk }$ and Rici
rotation coefficients
 $\gamma_{abc}$  are
\begin{eqnarray}
\Gamma^{r}_{\;\;jk } = \left | \begin{array}{ccc}
0 & 0 & \mbox{th}\;z \\
0 & - \mbox{sinh}\; r \; \mbox{cosh}\; r & 0 \\
\mbox{th}\;z & 0 & 0
\end{array} \right |  ,
\Gamma^{\phi}_{\;\;jk } = \left | \begin{array}{ccc}
0 & \mbox{cth}\; r & 0\\
\mbox{cth}\; r & 0 & \mbox{th}\; z \\
0 & \mbox{th}\; z & 0
\end{array} \right |  ,
\nonumber\\
\Gamma^{z}_{\;\;jk } = \left | \begin{array}{ccc}
-\mbox{cosh}\; z \;\mbox{sinh}\; z & 0 & 0\\
0 & -\mbox{sinh}\; z \; \mbox{cosh}\; z \; \mbox{sinh}^{2} r & 0 \\
0 & 0 & 0
\end{array} \right |  ,
\nonumber
\\
 \gamma_{12 2} =
 { 1 \over \mbox{cosh}\; z \mbox{tanh}\; r} \; , \qquad
 \gamma_{31 1} =
 \mbox{tanh}\; z\; , \qquad  \gamma_{32 2} =
 \mbox{tanh}\; z\; .
 \nonumber
 \label{A.4}
\end{eqnarray}

\noindent A general covariant Dirac equation (for more detail see
\cite{10})  takes the form
\begin{eqnarray}
 [ \;   i \gamma^{0} \; \partial_{t}  + {i \gamma^{1}  \over
\mbox {cosh}\, z}\;
 (    \partial_{r} + {1 \over 2}  {1 \over \mbox {th}\, r}  )
 + \gamma^{2} \;  {    i   \partial_{\phi} + e  B (\mbox {cosh}\,r -1)    \over \mbox {cosh}\, z\; \mbox {sinh}\, r}
 \nonumber
 \\
   +
 i \gamma^{3} \; ( \partial_{z} + \mbox {th}\, z   ) - M  \;   ] \;  \Psi = 0\; .
\label{A.6}
\end{eqnarray}

\noindent With the  substitution $\Psi = \psi /  \mbox {cosh}\,z \;  \sqrt{\mbox{sinh}\, r}
$ eq. (\ref{A.6}) becomes simpler
\begin{eqnarray}
\left [    i \gamma^{1}  \partial _{ r }  +
\gamma^{2} {    i   \partial_{\phi} + e  B (\mbox {cosh}\,r -1)
\over  \mbox {sinh}\, r}   +
  \mbox {cosh}\, z   (  i  \gamma^{0} \partial _{ t}  +
  i  \gamma^{3}  \partial _{ z} -  M    )  \right ] \psi =0\,.
\label{A.7}
\end{eqnarray}

\noindent Solutions of this equation will be searched in the form
\begin{eqnarray}
\psi = e^{-i\epsilon t} e^{im \phi} \left | \begin{array}{c}
f_{1}(r,z)\\
f_{2}(r,z)\\
f_{3}(r,z)\\
f_{4}(r,z)
\end{array} \right | ,
\nonumber
\end{eqnarray}

\noindent so that
\begin{eqnarray}
\left [  i \gamma^{1}  \partial _{r }  -  \gamma^{2} \mu (r)  +
  \mbox {cosh}\, z  \; ( \epsilon   \gamma^{0}   +
  i  \gamma^{3}  \partial_{ z} -  M    )  \right ]  \left | \begin{array}{c}
f_{1}(r,z)\\ f_{2}(r,z)\\
f_{3}(r,z)\\
f_{4}(r,z) \
\end{array} \right |
 =0  ,
\nonumber
 \label{A.8}
\end{eqnarray}

\noindent where
\begin{eqnarray}
 \mu (r) = {m -  e B (\mbox
{cosh}\, r -1)  \over  \mbox{sinh}\,  r } \;. \nonumber
\end{eqnarray}

\noindent  Taking the  Dirac matrices in spinor basis, we get
radial equations for  $f_{a}(t,z)$
\begin{eqnarray}
  ( \partial _{ r }  +   \mu  ) \; f_{4}
   +  \mbox {cosh} \;  z  \;   \partial_{z} \; f_{3}
+  i \; \mbox {cosh} \; z    \; ( \epsilon   f_{3}   -  M  f_{1} )
=0\, ,
\nonumber\\
( \partial_{ r }  -  \mu  ) \;f_{3} -  \mbox {cosh} \;
z\;    \partial_{z} \; f_{4}
 +   i  \; \mbox {cosh} \; z  \; (  \epsilon  f_{4}   - M  f_{2} ) =0\, ,
 \nonumber\\
 ( \partial_{r }  +    \mu )  \; f_{2}
+ \mbox {cosh}  \; z\;   \partial_{z} \; f_{1}
 - i   \; \mbox {cosh} \; z  \; ( \epsilon f_{1} -M f_{3} ) =0\, ,
 \nonumber\\
  ( \partial _{r }  -    \mu  ) \; f_{1}
-  \mbox {cosh}   \;z \;   \partial_{z} \; f_{2}  - i \;
\mbox {cosh} \; z  \; (   \epsilon f_{2} -M f_{4} ) =0\,  .
 \label{A.9}
\end{eqnarray}

To simplify the system (\ref{A.9}) one additional  operator should
be diagonalized. In flat space, as  that can be  taken the
helicity operator
\begin{eqnarray}
(\vec{\Sigma} \;\vec{P}) \; \Psi_{cart}  = \lambda \; \Psi_{cart}
\; , \;\; \vec{P} = - i \vec{\nabla} \;  . \nonumber
\end{eqnarray}

\noindent Because the cartesian and cylindrical bases are related
by spinor gauge transformation over fermion wave functions
\begin{eqnarray}
\Psi^{cyl} = S\; \Psi^{cart} \; , \;\; S = \left |
\begin{array}{cc} B  &  0  \\  0  &  B  \end{array} \right | \; ,
\;\; B =  \left | \begin{array}{cc} e^{+i\phi/2} & 0 \\ 0  &
e^{-i\phi/2}
 \end{array} \right | \; ,
\label{6.5a}
\end{eqnarray}

\noindent for the helicity operator in cylindrical representation
\begin{eqnarray}
B \;\vec{\Sigma} \;\vec{P} \; B^{-1}   = \left | \begin{array}{cc}
P_{3}  &   e^{+i\phi/2} \; ( P_{1} - iP_{2} ) \; e^{-i\phi/2}  \\
e^{+i\phi/2} \; ( P_{1} + iP_{2} ) \; e^{-i\phi/2}     & -P_{3}
\end{array} \right |   \; ,
\nonumber
\end{eqnarray}
where
\begin{eqnarray}
( P_{1} \pm i P_{2} ) = -i e^{\pm i\phi/2}
 ( {\partial \over \partial r }  \pm  {i \over r}
{\partial \over  \partial \phi}  ) \; , \nonumber
\\
e^{\mp i\phi/2} ( P_{1} \pm i P_{2} ) e^{\pm i\phi/2}  = [ -i  (
\partial _{ r} + {1 \over 2r} ) \pm {1 \over r }
\partial _{ \phi}  ] \; ,
\nonumber
\end{eqnarray}

\noindent one  produces the expression
\begin{eqnarray}
B \;\vec{\Sigma} \;\vec{P} \; B^{-1}    =
\gamma^{2}\gamma^{3}\; (  \partial _{r}   +  {1 \over 2r} ) \;   -i \gamma^{3}\gamma^{1} \;{i\partial _{\phi }
\over r} \;   +  \gamma^{1}
\gamma^{2} \;  \partial _{ z} \;   . \label{6.6}
\end{eqnarray}

\noindent
By the use of the substitution $ \Psi^{0} = \psi^{0} / \sqrt{r}$,
the operator  (\ref{6.6})  is simplified to
\begin{eqnarray}
\Sigma ^{0} = B \;\vec{\Sigma} \;\vec{P} \; B^{-1}    =    \gamma^{2}\gamma^{3}\; \partial _{ r }   \;
-i \gamma^{3}\gamma^{1} \;{i\partial _{\phi} \over r}
\; + \; \gamma^{1} \gamma^{2} \; \partial _{ z} \; . \label{6.6'}
\end{eqnarray}

In presence of external magnetic field in flat space, the Dirac equation looks
\begin{eqnarray}
 \left (    i \gamma^{0} \; \partial_{t}  + i \gamma^{1}
     \partial_{r}
 + \gamma^{2} \;  {    i   \partial_{\phi} + e  Br^{2} /2     \over
 r }+
 i \gamma^{3} \;  \partial_{z}  - M  \right    )   \psi^{0} = 0\; ,
\label{A.6''}
\end{eqnarray}

\noindent  hence an extended helicity operator is
\begin{eqnarray}
\Sigma_{0}   =    \gamma^{2}\gamma^{3}\; \partial _{ r }   \;
-i \gamma^{3}\gamma^{1} \;{i\partial _{\phi}  + e  Br^{2} /2  \over r}
\; + \; \gamma^{1} \gamma^{2} \; \partial _{ z} \; .
\end{eqnarray}

\noindent
We omit all details of calculation proving this.

In the Lobachevsky space, taking into consideration the explicit form of  the Dirac
equation
\begin{eqnarray}
 \left [    i \gamma^{0} \; \partial_{t}  + {1 \over
\mbox {cosh}\, z} \left (  i \gamma^{1}
     \partial_{r}
 + \gamma^{2} \;  {    i   \partial_{\phi} + e  B (\mbox {cosh}\,r -1)    \over
 \mbox {sinh}\, r} \right )
 + i \gamma^{3} \;  \partial_{z}  - M  \right    ]   \psi = 0\; ,
\nonumber
\label{A.6''}
\end{eqnarray}

\noindent
 one  may guess the form
of a generalized helicity operator $\Sigma$ in the curved space
\begin{eqnarray}
\Sigma =   {1 \over \mbox{cosh} z} \left (  \gamma^{2}\gamma^{3}\;  \partial _{r}    -i \;
\gamma^{3} \gamma^{1}  {i\partial _{\phi}  + e  B (\mbox {cosh}\,r -1)  \over \mbox {sinh}\, r } \right )
  +  \gamma^{1} \gamma^{2}  \partial _{z}    \; .
  \label{6.7}
\end{eqnarray}

Let us prove the commutation relation by direct calculation. It is convenient to make calculation in several steps.
Let it be
\begin{eqnarray}
\Sigma_{1}={1 \over \mbox{cosh} z} \gamma^{2}\gamma^{3}\;  \partial _{r}\,,
\qquad
[H,\Lambda_{1}]=H\Lambda_{1}-\Lambda_{1}H
\nonumber
\\
=i \gamma^{0} \; \partial_{t}\,{1 \over \mbox{cosh} z} \gamma^{2}\gamma^{3}\;  \partial _{r}
+{1 \over
\mbox {cosh}\, z} \, i \gamma^{1}
     \partial_{r}\,{1 \over \mbox{cosh} z} \gamma^{2}\gamma^{3}\;  \partial _{r}
\nonumber
\\
-{1 \over
\mbox {cosh}\, z}
  \gamma^{2} \;  \mu  {1 \over \mbox{cosh} z} \gamma^{2}\gamma^{3}\;  \partial _{r}+
i \gamma^{3} \; \partial_{z}\,{1 \over \mbox{cosh} z} \gamma^{2}\gamma^{3}\;  \partial _{r}
-M\,{1 \over \mbox{cosh} z} \gamma^{2}\gamma^{3}\;  \partial _{r}
\nonumber
\end{eqnarray}

\begin{eqnarray}
- {1 \over \mbox{cosh} z}   \gamma^{2}\gamma^{3}\;  \partial _{r}\, i \gamma^{0} \; \partial_{t}   - {1 \over \mbox{cosh} z}   \gamma^{2}\gamma^{3}\;  \partial _{r}\,{1 \over
\mbox {cosh}\, z} \,  i \gamma^{1}
     \partial_{r}
\nonumber
\\
+ {1 \over \mbox{cosh} z}   \gamma^{2}\gamma^{3}\;  \partial _{r}\,{1 \over
\mbox {cosh}\, z} \,\gamma^{2} \;  \mu
- {1 \over \mbox{cosh} z}   \gamma^{2}\gamma^{3}\;  \partial _{r}\,i \gamma^{3} \;  \partial_{z} +
{1 \over \mbox{cosh} z}   \gamma^{2}\gamma^{3}\;  \partial _{r}\, M=
 \nonumber
 \end{eqnarray}

 \begin{eqnarray}
 = - {1 \over
\mbox {cosh}^{2}\, z}
  \gamma^{2} \;  \mu  \gamma^{2}\gamma^{3}\;  \partial _{r}
 +
  {1 \over \mbox{cosh}^{2} z}
  \gamma^{2}\gamma^{3}\;  \partial _{r}\, \gamma^{2} \; \mu
  \nonumber
\\
+i \gamma^{3} \; \partial_{z}\,{1 \over \mbox{cosh} z} \gamma^{2}\gamma^{3}\;
   \partial _{r}- {1 \over \mbox{cosh} z}   \gamma^{2}\gamma^{3}\;  \partial _{r}\,i \gamma^{3} \;  \partial_{z}\,,
\nonumber
\end{eqnarray}

\noindent
that is
\begin{eqnarray}
[H,\Sigma_{1}] =
  \gamma^{3} {1 \over
   \mbox {cosh}^{2}\, z}  \mu  \partial _{r}
 +
 {1 \over \mbox{cosh}^{2} z}
  \gamma^{3}\;  (\partial _{r}   \mu)
  +i \gamma^{2} \; (\partial_{z}\,{1 \over \mbox{cosh} z}) \;
   \partial _{r}+
   i \gamma^{2} {1 \over \mbox{cosh} z}     \partial _{r} \;  \partial_{z}\,.
\nonumber
\end{eqnarray}

\noindent
Now, let it be
\begin{eqnarray}
\Sigma_{2}= {1 \over \mbox{cosh} z}\,i \;
\gamma^{3} \gamma^{1}  \mu (r)  \,, \qquad
[H,\Lambda_{2}]=H\Lambda_{2}-\Lambda_{2}H
\nonumber
\\
=
i \gamma^{0} \; \partial_{t}\,{1 \over \mbox{cosh} z}\,i \;
\gamma^{3} \gamma^{1}  \mu
+ {1 \over
\mbox {cosh}\, z}   i \gamma^{1}
     \partial_{r}\,{1 \over \mbox{cosh} z}\,i \;
\gamma^{3} \gamma^{1}  \mu
\nonumber
\\
- i \gamma^{2} \;
  \gamma^{3} \gamma^{1} {1 \over
\mbox {cosh}^{2} z}
    \mu^{2}   +i \gamma^{3} \; \partial_{z}\,{1 \over \mbox{cosh} z}\,i \;
\gamma^{3} \gamma^{1}  \mu
-M\,{1 \over \mbox{cosh} z}\,i \;
\gamma^{3} \gamma^{1}  \mu
\nonumber
\\
-{1 \over \mbox{cosh} z} \,i \;
\gamma^{3} \gamma^{1}  \mu  \, i \gamma^{0} \; \partial_{t} -
  {1 \over \mbox{cosh} z} \,i \;
\gamma^{3} \gamma^{1}  \mu  \, {1 \over
\mbox {cosh}\, z}  i \gamma^{1}
     \partial_{r}
 \nonumber
 \\
 +  {1 \over \mbox{cosh} z} \,i \;
\gamma^{3} \gamma^{1}  \, {1 \over
\mbox {cosh}\, z}
  \gamma^{2} \;  \mu^{2} (r)  - {1 \over \mbox{cosh} z} \,i \;
\gamma^{3} \gamma^{1}  \mu \, i \gamma^{3} \;  \partial_{z} +
 {1 \over \mbox{cosh} z} \,i \;
\gamma^{3} \gamma^{1}  \mu  \, M \; ,
\nonumber
\end{eqnarray}

\noindent
that is
\begin{eqnarray}
[H,\Sigma_{2}]= - {1 \over
\mbox {cosh}^{2} z}
     \gamma^{3}  (\partial_{r} \mu ) -   {1 \over \mbox{cosh}^{2} z}
\gamma^{3}   \mu        \partial_{r}
+
    \gamma^{1} (\partial_{z}\,{1 \over \mbox{cosh} z})  \mu
+ {1 \over \mbox{cosh} z} \, \gamma^{1} \mu     \partial_{z}\,.
\nonumber
\end{eqnarray}

\noindent
Now let it be
\begin{eqnarray}
\Sigma_{3}=\gamma^{1} \gamma^{2}  \partial _{z} \; , \qquad
[H,\Lambda_{3}]=H\Lambda_{0}-\Lambda_{3}H
\nonumber
\\
=i \gamma^{0} \; \partial_{t}\,\gamma^{1} \gamma^{2}  \partial _{z}
+{1 \over
\mbox {cosh}\, z}  i \gamma^{1}
     \partial_{r}\gamma^{1} \gamma^{2}  \partial _{z}
\nonumber
\\
-{1 \over
\mbox {cosh}\, z}
  \gamma^{2} \;  \mu  \gamma^{1} \gamma^{2}  \partial _{z}
+i \gamma^{3} \; \partial_{z}\,\gamma^{1} \gamma^{2}  \partial _{z}-
M\,\gamma^{1} \gamma^{2}  \partial _{z}
\nonumber
\\
-\gamma^{1} \gamma^{2}  \partial _{z} \, i \gamma^{0} \; \partial_{t}  -\gamma^{1} \gamma^{2}  \partial _{z} \, {1 \over
\mbox {cosh}\, z} \, i \gamma^{1}
     \partial_{r}
\nonumber
\\
 +\gamma^{1} \gamma^{2}  \partial _{z} \, {1 \over
\mbox {cosh}\, z} \,\gamma^{2} \; \mu
 -\gamma^{1} \gamma^{2}  \partial _{z} \, i \gamma^{3} \;  \partial_{z}  +\gamma^{1} \gamma^{2}  \partial _{z} \, M\; ,
\nonumber
\end{eqnarray}

so that
\begin{eqnarray}
 [H,\Sigma_{3}] =
 -  i   \gamma^{2}  {1 \over
\mbox {cosh}\, z}     \partial_{r}   \partial _{z}
-   i \gamma^{2}  (\partial _{z} \, {1 \over
\mbox {cosh}\, z} ) \,      \partial_{r}
    - \gamma^{1}  {1 \over
\mbox {cosh}\, z}
   \;  \mu    \partial _{z}   -  \gamma^{1}  \partial _{z} \, {1 \over
\mbox {cosh}\, z}  \;  \mu
 \,.
\nonumber
\end{eqnarray}

Summing tree commutators, we get
a needed one
\begin{eqnarray}
[H,\Sigma]=
 \gamma^{3}\;  {1 \over
   \mbox {cosh}^{2}\, z}  \mu \partial _{r}
 +
 {1 \over \mbox{cosh}^{2} z}
  \gamma^{3}\;  (\partial _{r}   \mu)
   \nonumber
\\
 +i \gamma^{2} \; (\partial_{z}\,{1 \over \mbox{cosh} z}) \;
   \partial _{r}+
   i \gamma^{2} {1 \over \mbox{cosh} z}     \partial _{r} \;  \partial_{z}
\nonumber
\\
- {1 \over
\mbox {cosh}^{2} z}
     \gamma^{3}  (\partial_{r} \mu ) -  \gamma^{3}   \;  {1 \over \mbox{cosh}^{2} z}
\mu        \partial_{r}
\nonumber
\\
+
    \gamma^{1} (\partial_{z}\,{1 \over \mbox{cosh} z})  \mu
+ {1 \over \mbox{cosh} z} \, \gamma^{1} \mu     \partial_{z}
\nonumber
\\
-  i   \gamma^{2}  {1 \over
\mbox {cosh}\, z}     \partial_{r}   \partial _{z}
-   i \gamma^{2}  (\partial _{z} \, {1 \over
\mbox {cosh}\, z} ) \,      \partial_{r}
 \nonumber
\\
     - \gamma^{1}  {1 \over
\mbox {cosh}\, z}
   \;  \mu    \partial _{z}   -  \gamma^{1}  (\partial _{z} \, {1 \over
\mbox {cosh}\, z})  \;  \mu
 = 0 \;.
\nonumber
\end{eqnarray}

Let us consider an eigenvalue equation
$\Sigma \psi = \sigma  \psi$:
\begin{eqnarray}
\left (
{   \gamma^{2}\gamma^{3}\;  \partial _{r}    +i \;
\gamma^{3} \gamma^{1}  \mu(r) \over \mbox{cosh} z}
  +  \gamma^{1} \gamma^{2}  \partial _{z}   \right )
   \left | \begin{array}{l}
f_{1}\\ f_{2}\\
f_{3}\\
f_{4} \
\end{array} \right |=
\sigma
\left | \begin{array}{l}
f_{1}\\ f_{2}\\
f_{3}\\
f_{4} \
\end{array} \right | ,
 \end{eqnarray}

\noindent
it leads to
\begin{eqnarray}
  ( \partial _{ r }  +   \mu  ) \; f_{4}
   +  \mbox {cosh} \;  z  \;   \partial_{z} \; f_{3}
-  i \, \mbox {cosh} \, z    \, \sigma \, f_{3}
=0\, ,
\nonumber\\
( \partial_{ r }  -  \mu  ) \;f_{3} -  \mbox {cosh} \;
z\;    \partial_{z} \; f_{4}
 -   i  \, \mbox {cosh} \, z  \,\sigma \, f_{4}    =0\, ,
 \nonumber\\
 ( \partial_{r }  +    \mu )  \; f_{2}
+ \mbox {cosh}  \; z\;   \partial_{z} \; f_{1}
 - i   \,\mbox {cosh} \, z  \,\sigma \, f_{1} =0\, ,
 \nonumber\\
  ( \partial _{r }  -    \mu  ) \; f_{1}
-  \mbox {cosh}   \;z \;   \partial_{z} \; f_{2}  - i \,
\mbox {cosh} \, z  \,\sigma \, f_{2}=0\,  .
\end{eqnarray}

Considering them together with previously obtained system (\ref{A.9})
\begin{eqnarray}
  ( \partial _{ r }  +   \mu  ) \; f_{4}
   +  \mbox {cosh} \;  z  \;   \partial_{z} \; f_{3}
+  i \; \mbox {cosh} \; z    \; ( \epsilon   f_{3}   -  M  f_{1} )
=0\, ,
\nonumber\\
( \partial_{ r }  -  \mu  ) \;f_{3} -  \mbox {cosh} \;
z\;    \partial_{z} \; f_{4}
 +   i  \; \mbox {cosh} \; z  \; (  \epsilon  f_{4}   - M  f_{2} ) =0\, ,
 \nonumber\\
 ( \partial_{r }  +    \mu )  \; f_{2}
+ \mbox {cosh}  \; z\;   \partial_{z} \; f_{1}
 - i   \; \mbox {cosh} \; z  \; ( \epsilon f_{1} -M f_{3} ) =0\, ,
 \nonumber\\
  ( \partial _{r }  -    \mu  ) \; f_{1}
-  \mbox {cosh}   \;z \;   \partial_{z} \; f_{2}  - i \;
\mbox {cosh} \; z  \; (   \epsilon f_{2} -M f_{4} ) =0\,  ,
 \nonumber
 \label{A.9'}
\end{eqnarray}

\noindent we arrive at a system of linear algebraic equations
\begin{eqnarray}
 \sigma \, f_{3}
+  ( \epsilon   f_{3}   -  M  f_{1} )
=0\, ,\qquad
\sigma \, f_{4}
 +    (  \epsilon  f_{4}   - M  f_{2} ) =0\, ,
 \nonumber
 \\
 \sigma \, f_{1}
 -  ( \epsilon f_{1} -M f_{3} ) =0\, ,
 \qquad
\sigma \, f_{2}
   - (   \epsilon f_{2} -M f_{4} ) =0\,  .
 \label{A.9'}
\end{eqnarray}

It gives
\begin{eqnarray}
\sigma = \mp p \; , \qquad p = \sqrt{\epsilon^{2} - m^{2}} \;,\qquad
f_{3} = {\epsilon \pm p  \over M} f_{1} \; ,\;
 f_{4} = {\epsilon \pm p  \over M} f_{2} \; .
\end{eqnarray}

With these linear restrictions we get two more simple systems

\vspace{4mm}
 $
 \sigma = -p\; ,
 $
\begin{eqnarray}
 ( {\partial \over \partial r }  +   \; \mu )  \; f_{2}
+ \mbox {cosh}\, z \;  ( { \partial  \over  \partial z}
 +  i  p   \; )\;   f_{1} =0\,,
 \nonumber\\
  ( {\partial \over \partial r }  -  \; \mu  ) \; f_{1}
-  \mbox {cosh}\, z  \; ({ \partial  \over  \partial z} - i  \; p\;
)\; f_{2} =0\, ; \label{A.14}
\end{eqnarray}

 $
 \sigma = +p\; ,
 $
\begin{eqnarray}
 ( {\partial \over \partial r }  +   \; \mu )  \; f_{2}
+ \mbox {cosh}\, z \;  ( { \partial  \over  \partial z}
 -  i  p\;)   \;    f_{1} =0\,,
  \nonumber\\
  ( {\partial \over \partial r }  -  \; \mu  ) \; f_{1}
-  \mbox {cosh}\, z  \; ( { \partial  \over  \partial z} + i  \; p\;
) \; f_{2} =0\, . \label{A.15}
\end{eqnarray}

For definiteness, let us consider the system   (\ref{A.14})
(transition to the case
 (\ref{A.15}) is performed by the formal change $p \Longrightarrow -p$).
 Let us search solutions in the form
\begin{eqnarray}
f_{1} = Z_{1} (z) \; R_{1} (r) \; ,  \qquad  f_{2} = Z_{2}(z) \;
R_{2} (r) \; . \label{A.16}
\end{eqnarray}

\noindent Eqs.   (\ref{A.14})  result in
\begin{eqnarray}
 ( {\partial \over \partial r }  +   \; \mu )  \; Z_{2} R_{2}
+ \mbox {cosh}\, z \;  ( { \partial  \over  \partial z}
 +  i  p   \; )\;   Z_{1}  R_{1} =0\,,
\nonumber\\
  ( {\partial \over \partial r }  -  \; \mu  ) \; Z_{1} R_{1}
-  \mbox {cosh}\, z  \; ({ \partial  \over  \partial z} - i  \; p\;
)\; Z_{2}  R_{2} =0\, . \label{A.17}
\end{eqnarray}

\noindent Introducing the separating constant  $\lambda$, we arrive at two   systems, in
variables  $z$ and $r$ respectively:
\begin{eqnarray}
 \mbox {cosh}\, z  \; ({  d  \over  d  z} + i  \; p\; )\;
Z_{1}  = \lambda  \; Z_{2} \; ,\qquad \mbox {cosh}\, z  \; ({
d  \over  d  z} - i  \; p\; )\; Z_{2} = \lambda \;
Z_{1}  \; ;  \label{A.18}
\end{eqnarray}
\begin{eqnarray} ( {d \over d r }  +   \; \mu )  \;  R_{2} +
\lambda  \; R_{1}  =0\,,\qquad
   ( { d  \over d r }  -  \; \mu  ) \;  R_{1}
-  \lambda   \; R_{2}  =0\, . \label{A.19}
\end{eqnarray}

\section{ Solution of the  differential equations in  $z$-variable}

\vspace{3mm}

From (\ref{A.18}) it follows  second  order differential equations
for
 $Z_{1}(z)$ and  $Z_{2}(z)$
\begin{eqnarray}
{d^{2}Z_{1} \over dz} + {\mbox {sinh}\, z  \over \mbox {cosh}\, z}
 {dZ_{1}\over dz} +  \left(   p^{2}+ip{\mbox {sinh}\, z\over \mbox {cosh}\,
z} -  {\lambda^{2}\over \mbox {cosh}\,^{2}z}
 \right)Z_{1}=0 \, ,
\label{A.20a}
\end{eqnarray}
\begin{eqnarray}
{d^{2}Z_{2} \over dz} + {\mbox {sinh}\, z  \over \mbox {cosh}\, z}
 {dZ_{2}\over dz} +  \left(   p^{2}-ip{\mbox {sinh}\, z\over \mbox {cosh}\,
z} -  {\lambda^{2}\over \mbox {cosh}\,^{2}z}
 \right)Z_{2}=0 \, .
\label{A.20b}
\end{eqnarray}

\noindent  Consider  the first equation for   $Z_{1}(z)$.   In a new variable  $  y = (1 + \mbox{tanh}\; z)/2$,
it  reads
\begin{eqnarray}
\left [ 4y (1-y)
 {d \over d y} +
2  (1-2y) {d \over d y}
\right.
\left. +
     p^{2} ( {1 \over 1- y}  + {1 \over y} ) +ip (  {1 \over 1-y } - {1 \over y}  )
-  4 \lambda^{2}
 \right ]  Z_{1}=0  \; .
\nonumber
\label{A.21}
\end{eqnarray}

\noindent With the substitution  $Z_{1} = y^{A} (1-y)^{B} \bar{Z}_{1} (y)$,
it  leads to
\begin{eqnarray}
4y\,(1-y)\,{d^{2} \bar{Z}_{1} \over dy^{2}}+4\left ( 2A\,+\,{1\over
2}-(2A\,+\,2B\,+1)\,y\right ){d \bar{Z}_{1}\over dy}
\nonumber \\
+[\; {2A\,(2A-1)+p\,(p-i)\over y}+{2B\,(2B-1)+p\,(p+i)\over
1-y}
\nonumber
\\
-4\,(A+B)^{2}
 -4\lambda^{2} \; ] \; \bar{Z}_{1}=0\,. \nonumber
\end{eqnarray}

\noindent With restrictions
\begin{eqnarray}
 A= -{ ip \over 2} \; ,\; {1 + ip \over 2} \, ,  \qquad B = {ip\over
2}\, ,\;{1 -ip \over 2} \, , \label{A.23}
\end{eqnarray}

\noindent the equation for  $\bar{Z}_{1}$ becomes   of
hypergeometric   type \cite{Bateman}
\begin{eqnarray}
y\,(1-y)\,{d^{2} \bar{Z}_{1}\over dz^{2}}+  [ \; 2A\,+\,{1\over
2}-(2A\,+\,2B\,+1)\,y \; ]{d \bar{Z}_{1}\over
dz}
-
\left[(A+B)^{2}+\lambda^{2}\right] \bar{Z}_{1} =0\,  \label{A.24}
\end{eqnarray}

\noindent with parameters given by
\begin{eqnarray}
a =  A + B  +  i\lambda \;, \qquad b = A+B -i\lambda \;,
 \nonumber
 \\
 c = 2A + {1\over 2} \; , \qquad Z_{1} =  y^{A} (1 - y ) ^{B}  \;
F(a, \; b, \; c;\;  y )  \; . \label{A.25}
\end{eqnarray}

For definiteness, let us choose
\begin{eqnarray}
\qquad A = {1 +i p \over 2} \;,\qquad B = {i p \over 2} \;, \qquad A+B = i p +1/2 \;  ;
\nonumber
\\
a = i p +1/2 + i \lambda \;, \qquad  b = i p +1/2 - i \lambda \;, \qquad  c = ip +3/2 \; ,
\end{eqnarray}

Two linearly independent solutions are
\begin{eqnarray}
Z_{1}^{(1)}  = y^{A} (1-y)^{B} U_{1}(y) = y^{A} (1-y)^{B} F(a,b,c,y)
\nonumber
\\
=
y^{+ip/2+1/2} (1-y)^{+ip/2}
F(a, \;  b, \; c;\;  y )
\nonumber
\\
=
y^{+ip/2+1/2} (1-y)^{-ip/2+1/2}
F(c-a, \;  c-b, \; c;\;  y  )
 \; ,
\label{A.25}
\end{eqnarray}
\begin{eqnarray}
Z_{1}^{(5)}  = y^{A} (1-y)^{B} U_{5}(y)
\nonumber
\\
= y^{-ip/2} (1-y) ^{+ip/2} F(a+1-c, b+1-c, 2-c,  y)
\nonumber
\\
= y^{-ip/2} (1-y) ^{-ip/2+1/2} F(1-a, 1- b, 2-c,  y) \; .
 \label{A.26}
\end{eqnarray}

At $z \rightarrow - \infty \;(y \rightarrow 0)$ they behave as follows
\begin{eqnarray}
Z_{1}^{(1)}  =
y^{+ip/2+1/2} \longrightarrow 0
  \; , \qquad
Z_{1}^{(5)}  = y^{-ip/2}\; .
 \label{A.27}
\end{eqnarray}

To  find behavior of these solutions  in the region $z  \rightarrow + \infty \; (y \rightarrow 1)$
one should make use of Kummer's relations
\begin{eqnarray}
U_{1} = {\Gamma(c) \Gamma (c-a-b) \over \Gamma (c-a) \Gamma (c-b)} \; U_{2} +
{\Gamma(c) \Gamma (a+b-c) \over \Gamma (a) \Gamma (b)} \; U_{6}  \; ,
\nonumber
\\
U_{5} = {\Gamma(2-c) \Gamma (c-a-b) \over \Gamma (1-a) \Gamma (1-b)} \; U_{2} +
{\Gamma(2-c) \Gamma (a+b-c) \over \Gamma (a+1-c) \Gamma (b+1-c)} \; U_{6}  \; ,
\label{6.1}
\end{eqnarray}

\noindent where two couples of linearly independent solutions are involved
\begin{eqnarray}
U_{1} (y) = F(a,b,c; y)  \;,
\nonumber
\\
 U_{5} = y^{1-c} F(a+1-c, b+1-c, 2-c, y )   \; ;
\nonumber
\\[3mm]
U_{2}(y) = F(a,b, a+b -c+1 ; 1-y)\;,
\nonumber
\\
U_{6} (y) = (1-y) ^{c-a-b} F(c-a, c-b, c-a-b +1; 1-y) \; .
\label{6.3}
\end{eqnarray}

Thus, for $Z_{1}^{(1)}$ we get the following  expansions
\begin{eqnarray}
Z_{1}^{(1)} =  { \Gamma(c) \Gamma (c-a-b) \over \Gamma (c-a) \Gamma (c-b)}
\nonumber
\\
\times
y^{+ip/2+1/2} (1-y)^{+ip/2}
\; F(a,b, a+b -c+1 ; 1-y)
\nonumber
\\
+ {\Gamma(c) \Gamma (a+b-c) \over \Gamma (a) \Gamma (b)} \;
y^{+ip/2+1/2} (1-y)^{+ip/2}
\nonumber
\\
\times
(1-y) ^{c-a-b} F(c-a, c-b, c-a-b +1; 1-y)  \; ,
\end{eqnarray}

Allowing for identity $c-a-b =  - ip - 1/2$,  we derive asymptotic  formula at $z \rightarrow + \infty \;(y \rightarrow 1)$
\begin{eqnarray}
Z_{1}^{(1)} \sim
{ \Gamma(c) \Gamma (c-a-b) \over \Gamma (c-a) \Gamma (c-b)} (1-y)^{+ip/2}
 + {\Gamma(c) \Gamma (a+b-c) \over \Gamma (a) \Gamma (b)} \;
(1-y) ^{-ip/2+1/2}  .
\nonumber
\\
\end{eqnarray}

Similarly, for $Z_{1}^{(5)}$ we produce
\begin{eqnarray}
Z_{1}^{(5)} = {\Gamma(2-c) \Gamma (c-a-b) \over \Gamma (1-a) \Gamma (1-b)}
\nonumber
\\
\times
 y^{-ip/2}
(1-y)^{+ip/2}\times F(a,b, a+b -c+1 ; 1-y)
\nonumber
\\
+ {\Gamma(2-c) \Gamma (a+b-c) \over \Gamma (a+1-c) \Gamma (b+1-c)}
\nonumber
\\
\times
y^{-ip/2}
(1-y)^{+ip/2}
(1-y) ^{c-a-b} F(c-a, c-b, c-a-b +1; 1-y) \; .
\nonumber
\\
\end{eqnarray}

\noindent
Allowing for identity $c-a-b =  - ip - 1/2$,
we derive an asymptotic  formula at $z \rightarrow + \infty \;  (y \rightarrow 1)$
\begin{eqnarray}
Z_{1}^{(5)} \sim
{\Gamma(2-c) \Gamma (c-a-b) \over \Gamma (1-a) \Gamma (1-b)}
 (1-y)^{+ip/2}
\nonumber
\\
+ {\Gamma(2-c) \Gamma (a+b-c) \over \Gamma (a+1-c) \Gamma (b+1-c)}
(1-y)^{-ip/2 +1/2} \; .
\end{eqnarray}

In the same manner, let us consider  the  equation for   $Z_{2}(z)$.
In the  new variable  $  y = (1 + \mbox{tanh}\; z)/2$,
it  reads
\begin{eqnarray}
\left [ 4y (1-y)
 {d \over d y} +
2  (1-2y) {d \over d y}   \right.
\left. +
     p^{2} ( {1 \over 1- y}  + {1 \over y} ) -ip (  {1 \over 1-y } - {1 \over y}  )
-  4 \lambda^{2}
 \right ]  Z_{2}=0 \; .
\nonumber
\label{A.21'}
\end{eqnarray}

\noindent With the substitution  $Z_{2} = y^{K} (1-y)^{L} \bar{Z}_{2} (y)$,
it  leads to
\begin{eqnarray}
4y\,(1-y)\,{d^{2}\bar{Z}_{2} \over dy^{2}}+4\left[2K\,+\,{1\over
2}-(2K\,+\,2L\,+1)\,y\right]{d\bar{Z}_{2}\over dy}
\nonumber \\
+[\; {2K\,(2K-1) -p\,(-p-i)\over y}+{2L\,(2L-1)-p\,(-p+i)\over
1-y}
\nonumber
\\
-4\,(K+L)^{2}
 -4\lambda^{2} \; ] \; \bar{Z}_{2} =0\,. \nonumber
\end{eqnarray}

\noindent With restrictions
\begin{eqnarray}
 K= { ip \over 2} \; ,\; {1 - ip \over 2} \, ,  \qquad L = -{ip\over
2}\, ,\;{1 +ip \over 2} \, , \label{A.23'}
\end{eqnarray}

\noindent the equation for  $\bar{Z}$ is reduced to that of
hypergeometric   type \cite{Bateman}
\begin{eqnarray}
y\,(1-y)\,{d^{2}\bar{Z}_{2} \over dz^{2}}+  [ \; 2K\,+\,{1\over
2}-(2K\,+\,2L\,+1)\,y \; ]{d\bar{Z}_{2}\over
dz}
\nonumber
\\
-
\left[(K+L)^{2}+\lambda^{2}\right]\bar{Z}_{2} =0\,  \label{A.24'}
\end{eqnarray}

\noindent with parameters given by
\begin{eqnarray}
\alpha  =  K + L  +  i\lambda \;, \qquad \beta  = K + L  -i\lambda \;,
 \nonumber
 \\
 \gamma = 2K + {1\over 2} \; ,
 \qquad Z_{2} =  y^{K} (1-y) ^{L}  \;
F(\alpha, \; \beta, \; \gamma ;\;  y )  \; . \label{A.25}
\end{eqnarray}

For definiteness, let us choose
\begin{eqnarray}
\qquad K = { +i p \over 2} \;,\qquad  L = {i p +1 \over 2} \;, \qquad K+L = i p +1/2 \;  ,
\nonumber
\\
\alpha  = i p +1/2 + i \lambda \;, \qquad  \beta  = i p +1/2 - i \lambda \;, \qquad  \gamma  = ip +1/2 \; .
\end{eqnarray}

Two linearly independent solutions are
\begin{eqnarray}
Z_{2}^{(1)}  = y^{K} (1-y)^{L} U_{1}(y) = y^{K} (1-y)^{L} F(\alpha, \beta, \gamma, y)
\nonumber
\\
=
y^{+ip/2} (1-y)^{+ip/2+1/2}
F(\alpha, \;  \beta, \; \gamma;\; y  )  \; ,
\label{A.25'}
\end{eqnarray}
\begin{eqnarray}
Z_{2}^{(5)}  = y^{K} (1-y)^{L} U_{5}(y)
\nonumber
\\
=
 y^{K} (1-y)^{L} \; y^{1-\gamma} F(\alpha+1-\gamma, \beta+1-\gamma, 2-\gamma,y)
\nonumber
\\
= y^{-ip/2+1/2}
(1-y)^{+ip/2+1/2}
F(\alpha+1-\gamma, \beta+1-\gamma, 2-\gamma,  y)\;.
 \label{A.26}
\end{eqnarray}

Among the functions $Z_{1}^{(1)}(y),  Z_{1}^{(5)}(y), Z_{2}^{(1)}(y),Z_{2}^{(5)}(y)$
there must exist pairs related by the first order  system (see (\ref{A.18}))
\begin{eqnarray}
\left  ( y (1-y) {d \over dy} + {ip \over 2} \right ) Z_{1} =  \lambda  \sqrt{y(1-y)}   \; Z_{2} \; ,
\nonumber
\\
 \left  ( y (1-y) {d \over dy} - {ip \over 2} \right ) Z_{2} =  \lambda  \sqrt{y(1-y)}   \; Z_{1} \; .
\label{*}
\end{eqnarray}

Let us start with
\begin{eqnarray}
Z_{1}^{(1)}  =z_{1}^{(1)}
y^{+ip/2+1/2} (1-y)^{+ip/2} F(a, \;  b, \; c;\;  y ) \;,
\nonumber
\\
Z_{2}^{(1)}  = z_{2}^{(1)} y^{+ip/2} (1-y)^{+ip/2+1/2} F(\alpha, \;  \beta, \; \gamma;\; y  )  \;,
\nonumber
\\
a= \alpha = ip +1/2 + i \lambda\; ,
\;\;
b = \beta = ip +1/2 - i \lambda \; ,
\nonumber
\\
c = \gamma +1=ip+3/2 \; .
\end{eqnarray}

\noindent Substituting these expressions into the second equation
\begin{eqnarray}
\left  ( y (1-y) {d \over dy} - {ip \over 2} \right ) Z_{2}^{(1)} =  \lambda  \sqrt{y(1-y)}   \; Z_{1}^{(1)} \; ,
\nonumber
\end{eqnarray}

\noindent we get
\begin{eqnarray}
z_{2}^{(1)}\left  ( y (1-y) {d \over dy} - {ip \over 2} \right ) y^{+ip/2} (1-y)^{+ip/2+1/2}
F(\alpha,   \beta,  \gamma; y  )
\nonumber
\\
=  z_{1}^{(1)} \lambda  \;   y^{+ip/2+1} (1-y)^{+ip/2+1/2 } F(a,   b,  c;  y ) \; ,
\nonumber
\end{eqnarray}

\noindent
 or
 \begin{eqnarray}
z_{2}^{(1)}\left  ( +{ip\over 2}  (1-y) - ({ip\over 2} +{1 \over 2})y +y(1-y){d \over dy} - {ip\over 2}
   \right )  F(a, b,  c -1; y  )
\nonumber
\\
=  z_{1}^{(1)} \lambda  \;   y  F(a,   b,  c;  y ) \; .
\nonumber
\end{eqnarray}

\noindent
From whence it follows
\begin{eqnarray}
z_{2}^{(1)}\left  (
-ip - {1 \over 2} + (1-y){d \over dy}
   \right )  F(a, b,  c -1; y  ) =
  z_{1}^{(1)} \lambda  \;     F(a,   b,  c;  y ) \; ,
\nonumber
\end{eqnarray}

\noindent or
 differently
 \begin{eqnarray}
z_{2}^{(1)}\left  (
1-c  + (1-y){d \over dy}
   \right )  F(a, b,  c -1; y  ) =
  z_{1}^{(1)} \lambda  \;     F(a,   b,  c;  y ) \; .
\label{49}
\end{eqnarray}

There  exist a  formula, relating derivatives of hypergeometric functions with
contiguous hypergeometric functions
\begin{eqnarray}
\left  (
1-c  + (1-y){d \over dy}
   \right )  F(a, b,  c -1, y  )
     =
   {(a-c+1)(b-c+1)\over c-1}  F(a,b,c,y) \;.
\nonumber
\label{50}
\end{eqnarray}

\noindent
Hence  relation (\ref{49}) gives
\begin{eqnarray}
z_{2}^{(1)} {(a-c+1)(b-c+1)\over c-1}  F(a,b,c,y)
 =
  z_{1}^{(1)} \lambda  \;     F(a,   b,  c;  y ) \; ,
\nonumber
\end{eqnarray}

\noindent from whence it follow
\begin{eqnarray}
z_{2}^{(1)} {(a-c+1)(b-c+1)\over c-1}
 =
  z_{1}^{(1)} \lambda  \;     \; .
\label{51}
\end{eqnarray}

Let us consider the second pair of solutions, starting   with
\begin{eqnarray}
Z_{1}^{(5)}
= y^{-ip/2} (1-y) ^{+ip/2} F(a+1-c, b+1-c, 2-c,  y) \; ,
\nonumber
\\
a = i p +1/2 + i \lambda \;, \qquad  b = i p +1/2 - i \lambda \;, \qquad  c = ip +3/2 \; ,
\nonumber
\end{eqnarray}
and
\begin{eqnarray}
Z_{2}^{(5)}  =   y^{-ip/2+1/2}
(1-y)^{+ip/2+1/2}
F(\alpha+1-\gamma, \beta+1-\gamma, 2-\gamma,  y)
 \nonumber
 \\
\alpha  = i p +1/2 + i \lambda \;, \qquad  \beta  = i p +1/2 - i \lambda \;, \qquad  \gamma  = ip +1/2 \; .
\nonumber
\end{eqnarray}

\noindent Firstly, it should be noted
\begin{eqnarray}
a' = a+1-c = + i \lambda\;, \qquad \alpha  +1 - \gamma =+ i \lambda + 1 = a' +1 \; ,
\nonumber
\\
b' = b+1-c = -i \lambda\;, \qquad \beta +1 - \gamma = i \lambda + 1 = b' +1 \; ,
\nonumber
\\
2-c =  -ip +1/2 = c' \;, \qquad 2 -\gamma =-ip +3.2 = c' +1 \; ,
\nonumber
\end{eqnarray}

\noindent Substituting the above expressions into the first equation
\begin{eqnarray}
z_{1}^{(5)}\left  ( y (1-y) {d \over dy} - {ip \over 2} \right ) y^{+ip/2} (1-y)^{+ip/2+1/2}
F(a',   b', c'; y  )
\nonumber
\\
=  z_{2}^{(5)} \lambda  \;   y^{+ip/2+1} (1-y)^{+ip/2+1/2 } F(a'+1,   b'+1,  c'+1;  y ) \; ,
\nonumber
\end{eqnarray}

\noindent we get
\begin{eqnarray}
z_{1}^{(5)} \left ( -{ip \over 2} (1-y) - {ip \over 2} y + y (1-y) {d \over dy}  +{ip\over 2} \right )
F(a',   b', c'; y  )
\nonumber
\\
=
z_{2}^{(5)} \lambda  \;  y(1-y) F(a'+1,   b'+1,  c'+1;  y ) \;.
\nonumber
\end{eqnarray}

\noindent It  reduces to
\begin{eqnarray}
z_{1}^{(5)}    y (1-y) {d \over dy}
F(a',   b', c'; y  ) =
z_{2}^{(5)} \lambda  \;  y(1-y) F(a'+1,   b'+1,  c'+1;  y ) \; ,
\nonumber
\end{eqnarray}

\noindent so we arrive at
\begin{eqnarray}
z_{1}^{(5)}    {a' b' \over c'} = z_{2}^{(5)} \lambda \; .
\end{eqnarray}

\section{ Solution of the equations in  $r$-variable}

From radial equations   (\ref{A.19}) it follows  second order
equations for $R_{1}$ and $ R_{2}$
\begin{eqnarray}
 ( {d^{2} \over dr^{2} } - {d \mu \over dr}  - \mu^{2} +
\lambda^{2}  )  \; R_{1} = 0 \; ,\qquad
 ( {d^{2} \over dr^{2} } + {d \mu \over dr}  - \mu^{2} +
\lambda^{2}  )  \; R_{2} = 0 \; .
\label{A.30}
\end{eqnarray}

\noindent Remembering on the meaning of  $\mu (r)$
  we obtain  their explicit t form  (for shortness let  us note  $eB$  as $B$)
\begin{eqnarray}
\left ( {d^{2}\over dr^{2}}+{m\,\mbox {cosh}\,  r+B\,(\mbox
{cosh}\,  r-1)\over \mbox {sinh}^{2}  r }-{[m-B\,(\mbox {cosh}\,
r-1)]^{2}\over \mbox {sinh}^{2}  r }+\lambda^{2}\right )R_{1}=0\,,
\nonumber
\\
\left( {d^{2} \over dr^{2}} - {m\,\mbox {cosh}\,  r+B\,(\mbox
{cosh}\,  r-1)\over \mbox {sinh}^{2}  r }-{[m-B\,(\mbox {cosh}\,
r-1)]^{2}\over \mbox {sinh}^{2}  r }+\lambda^{2}\right ) R_{2}=0\,.
\nonumber
\end{eqnarray}

\noindent Two equations are related by the formal change
\begin{eqnarray}
m \Longrightarrow - m\;, \qquad
B \Longrightarrow - B \;, \qquad R_{1} \Longrightarrow R_{2}\; .
\nonumber
\end{eqnarray}

\noindent
With the variable  $ y= (1+\mbox {cosh}\, r ) /2 $, making the substitution
\begin{eqnarray}
 R_{1} = y^{A} (1-y)^{C} \bar{R}_{1} (y)
\;,
\nonumber
\end{eqnarray}

\noindent
 one gets an  equation  for $\bar{R}_{1}$
\begin{eqnarray}
y(1-y){d^{2} \bar{R}_{1}\over dy^{2}}+ \left [  2A + {1\over 2}-(2A+2C +1)\;
y  \right] \,{d\bar{R}_{1}\over dy}   \nonumber
\\
+ \left[ \; { A^{2}-A/2-m^{2}/4-m/4- mB - B^{2}-B/2\over y}
\right. \nonumber
\\
\left.+ \; {C^{2}-C/2- m^{2}/4 + m/4\over 1-y}
-(A+C)^{2}-\lambda^{2}+B^{2} \; \right] \bar{R}_{1}=0\,. \label{A.28}
\end{eqnarray}

\noindent With restrictions
\begin{eqnarray}
 A= - {2B + m\over 2}\; ,\;  {2B + m +1 \over 2} \, , \qquad
 C = {m\over 2}\;  ,\; {1 -m \over 2} \; , \label{A.29}
\end{eqnarray}

\noindent we arrive at an equation of hypergeometric type
\begin{eqnarray}
y(1-y){d^{2}\bar{R}_{1}\over dy^{2}}+  [ 2A+{1\over 2}-(2A+2C+1)\;
y  ] \, {d\bar{R}_{1}\over dy}
\nonumber
\\
-   [ \; (A+C)^{2}+\lambda^{2}-B^{2}
\;  ] \; \bar{R}_{1} =0\,, \nonumber \label{A.30'}
\end{eqnarray}

\noindent so that
\begin{eqnarray}
R_{1} = y ^{A} \; (1 - y)^{C}
\; F(\alpha, \beta, \gamma ;\; y  ) \; , \qquad \gamma = 2A\,+\,{1\over 2}
\; ,
 \nonumber
\\
\alpha = A+C +\sqrt{B^{2}-\lambda^{2}}\;, \qquad
 \beta =
A+C-\sqrt{B^{2}-\lambda^{2}} \;.
 \label{A.32}
\end{eqnarray}

\noindent
To have solutions finite in the origin $r=0 \;(y \rightarrow 1)$ (corresponding
geometrical points belong to the axis  $z$: $ u_{0} =
\mbox{cosh}\; z  , \;  u_{3} = \mbox{sinh}\; z  , \; u_{1}=0 , \;
u_{2}=0$ ) and in infinity  $r \rightarrow \infty \; (y \rightarrow \infty)$, we must take
positive  $C$ and negative   $A$, such that
\begin{eqnarray}
  C >  0 \; , \qquad   \; A < 0 \;, \qquad C+A<0 \; ,
\nonumber
\\
R_{1} = y ^{A} \; (1 - y)^{C}
\; F(\alpha, \beta, \gamma ;\; y ) \;
 .
 \label{A.33}
\end{eqnarray}

Let us write down  all four possibilities to choose the parameters (for definiteness let it be  $B>0$)
\begin{eqnarray}
1.  \qquad \qquad C = {m \over 2}  \;  , \qquad  A = -
{2B + m  \over 2}  \; ,\qquad
  C+A = - B   \; ; \nonumber
\\
2. \qquad C = {1-m \over 2}  \; , \; A =  - {2B + m  \over
2} \; ,
\qquad   C+A = - B  - m  +{1 \over 2}   \; ;
\nonumber
\\
3. \qquad C = {m \over 2}  \;  , \; A = {2B + m +1 \over
2} \; ,
\qquad \;  C+A = B + m + {1 \over 2}   \; ;
 \nonumber
\\
4. \qquad C = {1-m \over 2}  \; , \; A = {2B + m +1
\over 2}  \; ,
\qquad
 C+A = B + 1  \; .
\nonumber
\\
 \label{A.33'}
\end{eqnarray}

Regarding (\ref{A.33}) we conclude that  only  variants 1 and 2 are acceptable
 for describing bound states (they coincide when $m=+1/2$)
\begin{eqnarray}
\underline{ \mbox{Variant  }} \;  1, \qquad   m >  0 \qquad  \qquad  (m = +1/2, +3/2, ...) \; ;
 \nonumber
 \\
\underline{ \mbox{Variant  }} \;  2, \qquad   -B +1/2  < m < 1 \qquad (m =  m_{min}, ...., -1/2 , +1/2 ) \;  .
\nonumber
\\
 \label{A.33''}
\end{eqnarray}

\noindent Respective expressions  for radial functions are

\vspace{2mm}
$
\underline{ \mbox{Variant  }} \; 1 \; , \; m =+1/2, +3/2, ...,$
\begin{eqnarray}
 C = m / 2 \;  , \qquad  A =  - B - m  / 2  < 0 \; ,\;
R_{1} = y ^{-B - m/2} \; (1 - y)^{m/2} \;  F(a, b, c ;\; y ) \; , \nonumber
\\
a = -B +\sqrt{B^{2}-\lambda^{2}}\;, \;\;  b = -B
-\sqrt{B^{2}-\lambda^{2}} \; ,
\qquad  c = -2B - m + {1\over 2}\; ;
\nonumber
 \label{A.34}
\end{eqnarray}

\noindent with the quantization rule
\begin{eqnarray}
a = -n \; \; \Longrightarrow \;\;  \;
\sqrt{B^{2}-\lambda^{2}} = B - n \; \; \Longrightarrow \;\;
\lambda^{2} = B^{2} - (B-n)^{2}  \; ;
 \label{A.35}
\end{eqnarray}

\noindent to have radial function finite at the infinity
 $r \rightarrow \infty $,  the following inequality  must be imposed
\begin{eqnarray}
A+ C + n < 0 \qquad \Longrightarrow \qquad    B -n > 0\; ;
 \label{A.36}
\end{eqnarray}

\noindent  which insures the positive square root
$+\sqrt{B^{2}-\lambda^{2}} $ in (\ref{A.35}).

\vspace{2mm}

$\underline{ \mbox{Variant  } \; 2} \; ,  \;\;  -B +1/2  < m < 1 \qquad (m =  m_{min}, ...., -1/2 , +1/2 ) \; ,$
\begin{eqnarray}
 C = 1/2 - m / 2 \;  , \qquad
 A =  - B - m  / 2  < 0 \; ,
\nonumber
\\
R_{1} = y ^{-B - m/2} \; (1 - y)^{1/2 - m/2} \;  F(a', b', c' ;\; y ) \; , \nonumber
\\
a' = - B  - m  + 1 / 2
 +\sqrt{B^{2}-\lambda^{2}}\;,
 \nonumber
 \\
 b' = - B  - m  + 1 / 2  -\sqrt{B^{2}-\lambda^{2}} \;,
 \;
 c' = -2B - m + {1\over 2}  \; ;
 \label{A.37}
\end{eqnarray}

\noindent the quantization rule is
\begin{eqnarray}
a' = - n \qquad \Longrightarrow \qquad
 + \sqrt{B^{2}-\lambda^{2}} =   B  + m  - 1 / 2 - n
 \nonumber
 \\
\Longrightarrow \qquad \lambda^{2} = B^{2} - ( B  +m -{1\over 2} -  n)  ^{2} \; ;
 \label{A.38}
\end{eqnarray}

\noindent an inequality must be fulfilled
\begin{eqnarray}
A+ C + n < 0 \qquad \Longrightarrow \qquad   B  + m  - 1 / 2  - n
> 0\; , \label{A.39}
\end{eqnarray}

\noindent  which make positive  the  root
$+\sqrt{B^{2}-\lambda^{2}} $ in  (\ref{A.38}).
Note that when $m = +1/2$, the formula (\ref{A.38}) give the same result as
 (\ref{A.35}).

Thus, the energy spectrum for spin $1/2$ particle in the
magnetic  field in the Lobachevsky space model is given by two
formulas
\begin{eqnarray}
\underline{ \mbox{Variant  }} \; 1.
  \qquad   \lambda^{2} = B^{2} - (B-n)^{2}  \;,\;\;
   \underline{m=+1/2, +3/2, ... \; , } \qquad  n < B \; ; \nonumber
\\[3mm]
\underline{ \mbox{Variant  }} \; 2. \qquad
\lambda^{2} = B^{2} - ( B  +m -{1\over 2} -  n)  ^{2} \;,
\nonumber
\\
\underline{-B+1/2  < m_{min} , ...,  +1/2  \;  ,}
 \qquad   n < B  + m  - 1 / 2 \; . \label{A.40}
\end{eqnarray}

Now we should construct explicit form of radial functions $R_{2}(y)$.
Analysis is similar (but with significant differences).
In the variable  $ y= (1+\mbox {cosh}\, r ) /2 $, making the substitution
\begin{eqnarray}
 R_{2} = y^{K} (1-y)^{L} \bar{R}_{2} (y)
\;,
\nonumber
\end{eqnarray}

\noindent
 one gets equation  $\bar{R}_{2}$
\begin{eqnarray}
y(1-y){d^{2} \bar{R}_{2}\over dy^{2}}+ \left [  2K + {1\over 2}-(2K+2L +1)\;
y  \right] \,{d\bar{R}_{2}\over dy} \;  \nonumber
\\
+ \left[ \; { K^{2}-K/2-m^{2}/4+m/4 - mB - B^{2}-B/2\over y} \;
\right. \nonumber
\\
\left.+ \; {L^{2}-L/2- m^{2}/4 - m/4\over 1-y}
-(K+L)^{2}-\lambda^{2}+B^{2} \; \right] \bar{R}_{2}=0\,. \label{A.28'}
\end{eqnarray}

\noindent With restrictions
\begin{eqnarray}
 K=  {2B + m\over 2}\; ,\;  {-2B - m +1 \over 2} \, , \qquad
 L = -{m\over 2}\;  ,\; {1 +m \over 2} \; , \label{A.29}
\end{eqnarray}

\noindent we arrive at an equation of hypergeometric type
\begin{eqnarray}
y(1-y){d^{2}\bar{R}_{2}\over dy^{2}}+  [ 2K+{1\over 2}-(2K+2L+1)\;
y  ] \, {d\bar{R}_{2}\over dy}
\nonumber
\\
-   [ \; (K+L)^{2}+\lambda^{2}-B^{2}
\;  ] \; \bar{R}_{2} =0\,, \nonumber \label{A.30''}
\end{eqnarray}

\noindent so that
\begin{eqnarray}
R_{2} = y ^{K} \; (1 - y)^{L}
\; F(\alpha', \beta', \gamma' ;\; y  ) \; , \qquad \gamma' = 2K\,+\,{1\over 2}
\; ,
 \nonumber
\\
\alpha' = K+L +\sqrt{B^{2}-\lambda^{2}}\;, \qquad
 \beta '=
K+L-\sqrt{B^{2}-\lambda^{2}} \;.
 \label{A.32'}
\end{eqnarray}

\noindent
To have  solutions finite at $r=0 \;(y \rightarrow 1)$
and at  $r \rightarrow \infty \; (y \rightarrow \infty)$, we must take
positive  $L$ and negative   $K$, such that
\begin{eqnarray}
  L >  0 \; , \qquad   \; K < 0 \;, \qquad L+K<0 \; ,
\nonumber
\\
R_{2} (y) = y ^{K} \; (1 - y)^{L}
\; F(\alpha', \beta', \gamma ';\; y ) \;
 .
 \label{A.33'}
\end{eqnarray}

Let us write down  all four possibilities to choose the parameters (remembering that  $B>0$)
\begin{eqnarray}
1'.  \qquad \qquad \qquad L = {-m \over 2}  \;  , \; K =
{2B + m  \over 2}  \; ,\qquad
  L+K =  B  \;\;\;  (NO)\; ; \nonumber
\\
2'. \qquad L = {1+m \over 2}  \; , \; K =   {2B + m  \over
2} \; ,
\qquad   L+K =  B  + m  +{1 \over 2}   \;\;\;   (NO)\; ;
\nonumber
\\
3'. \qquad L = -{m \over 2}  \;  , \; K = {-2B - m +1 \over
2} \; ,
\qquad \;  L+K = -B - m + {1 \over 2}   \; ;
 \nonumber
\\
4'. \qquad L = {1+m \over 2}  \; , \; K = {-2B - m +1
\over 2}  \; ,
\qquad
 L+K = -B + 1  \; .
 \label{A.33'}
\end{eqnarray}

Regarding (\ref{A.33}) we conclude that  only two variants, 3' and 4',  are acceptable
 for describing bound states
\begin{eqnarray}
\underline{ \mbox{Variant  }} \;
 4', \qquad   m > -1 \;, \;  \qquad  B> 1 \qquad  (m = -1/2, +1/2, +3/2, ...)  ;
 \nonumber
 \\
\underline{ \mbox{Variant  }} \;
 3', \qquad   -B +1/2 < m < 0  \qquad (m =  m_{min}, ...., -1/2  ) \;  .
 \label{A.33''}
\end{eqnarray}

\noindent Respective expressions  for radial functions are

\vspace{2mm}

$
\underline{ \mbox{Variant  }} \; 4'
 \; , \; m = \underline{-1/2}, +1/2, +3/2, ...,$
\begin{eqnarray}
 L = m/2 + 1/ 2  \;  , \qquad K =  - B - m  / 2  +1/2 < 0 \; ,
 \nonumber
 \\
  L+K = -B+1 <0\; ,
 \qquad  \gamma = -2B - m + {3\over 2}\; ,
\nonumber
\\
\alpha  = -B +1 +\sqrt{B^{2}-\lambda^{2}}\;, \;\;  \beta = -B+1  -\sqrt{B^{2}-\lambda^{2}} \; ,
\nonumber
\\
R_{2} = y ^{-B - m/2+1/2} \; (1 - y)^{m/2+1/2 } \;  F(\alpha, \beta, \gamma ;\; y ) \; ,
 \label{A.34'}
\end{eqnarray}

\noindent with the quantization rule
\begin{eqnarray}
\alpha = -N \; \; \Longrightarrow \;\;  \;
\sqrt{B^{2}-\lambda^{2}} = B - N-1 \; \; \Longrightarrow
\nonumber
\\
\lambda^{2} = B^{2} - (B-N-1)^{2}  \; ;
 \label{A.35'}
\end{eqnarray}

\noindent to have radial function finite at the infinity
 $r \rightarrow \infty $,  the following inequality  must be imposed
\begin{eqnarray}
K+ L + N < 0 \qquad \Longrightarrow \qquad    B -N -1 > 0\; ,
 \label{A.36'}
\end{eqnarray}

\noindent  which insures the positive square root
$+\sqrt{B^{2}-\lambda^{2}} $ in (\ref{A.35'}).

\vspace{4mm}

$\underline{ \mbox{Variant  } \; 3'} \; ,  \;\;  -B+1/2  < m < 0  \qquad (m =  m_{min}, ...., -1/2  ) \; ,$

\begin{eqnarray}
 L = - m / 2 \;  , \qquad
 K =  - B - m  / 2 +1/2  < 0 \; ,
 \nonumber
 \\
 \;\; K+L = -B - m +1/2\;, \qquad
 \gamma' = -2B - m +  {3\over 2}  \; ,
\nonumber
\\
\alpha' = - B  - m  + 1 / 2  +\sqrt{B^{2}-\lambda^{2}}\;,
 \nonumber
 \\
 \beta ' = - B  - m  + 1 / 2  -\sqrt{B^{2}-\lambda^{2}} \;,
\nonumber
\\
R_{2} = y ^{-B - m/2+1/2} \; (1 - y)^{ - m/2} \;  F(\alpha', \beta', \gamma' ;\; y ) \; ;
 \nonumber
\\
 \label{A.37'}
\end{eqnarray}

\noindent the quantization rule is
\begin{eqnarray}
\alpha' = - N \qquad \Longrightarrow \qquad
 + \sqrt{B^{2}-\lambda^{2}} =   B  + m  - 1 / 2 - n
 \nonumber
 \\
\Longrightarrow \qquad \lambda^{2} = B^{2} - ( B  +m -{1\over 2} -  n)  ^{2} \; ;
 \label{A.38'}
\end{eqnarray}

\noindent the inequality must be fulfilled
\begin{eqnarray}
A+ C + n < 0 \qquad \Longrightarrow \qquad   B  + m  - 1 / 2  - n
> 0\; , \label{A.39'}
\end{eqnarray}

\noindent  which make positive  the  root
$+\sqrt{B^{2}-\lambda^{2}} $ in  (\ref{A.38'}).

Analysis of the bound states from equation for $R_{2}$ can be summarized by the formula

Let us collect results on energy level obtained from analysis of $R_{1}(y)$ and $R_{2}(y)$:
\begin{eqnarray}
\underline{R_{1}(y) }\;,  \qquad 1.
  \qquad   \lambda^{2} = B^{2} - (B-n)^{2}  \;,
  \nonumber
  \\
   m=+1/2, +3/2, ... \; ,  \qquad  n < B \; ; \nonumber
\\[3mm]
2. \qquad
\lambda^{2} = B^{2} - ( B  +m -{1\over 2} -  n)  ^{2} \;,
\nonumber
\\
-B +1/2 < m = m_{min} , ..., \underline{-1/2},  +1/2  \;  ,
\nonumber
\\
  n < B  + m  - 1 / 2 \; ; \label{A.40}
\end{eqnarray}

\noindent
and
\begin{eqnarray}
\underline{R_{2}(y)}\; ,   \qquad 4'. \qquad
\lambda^{2} = B^{2} - (B-N-1)^{2} \; ,
\nonumber
\\
 m = \underline{-1/2}, +1/2, +3/2, ... ; \qquad
 N +1  < B    \; ;
 \nonumber
 \end{eqnarray}
 \begin{eqnarray}
 3'. \qquad
 \lambda^{2} = B^{2} - ( B  +m -{1\over 2} -  n)  ^{2} \; ,
 \nonumber
 \\
-B +1/2 < m = m_{min},  ...., \underline{-1/2 }\;  ,
\nonumber
\\
 n < B  + m  - 1 / 2  \; .
\end{eqnarray}

We see evident correlations between the cases  1 and $4'$, as well as between the cases 2 and $3'$.
To detail them, let us turn to the system  (\ref{A.19}) relating functions $R_{1}$ and $R_{2}$
\begin{eqnarray} ( {d \over d r }  +   \; \mu )  \;  R_{2} +
\lambda  \; R_{1}  =0\,,
\qquad
   ( { d  \over d r }  -  \; \mu  ) \;  R_{1}
-  \lambda   \; R_{2}  =0\, . \nonumber
\label{A.19'}
\end{eqnarray}

\noindent
After translating to the variable $y=(1 +\mbox{cosh}\; r )/2 $
\begin{eqnarray}
 {d \over dr} = \sqrt{-y(1-y)}{d \over dr}\;,\qquad
\mbox{sinh}\; r = 2 \sqrt{-y(1-y)}\;,
\nonumber
\\
 \mu =
{m - B (\mbox{cosh}\; r -1)\over \mbox{sinh}\; r } ={ m -2B (y-1) \over 2 \sqrt{-y(1-y)}}\; ,
\label{A.20}
\end{eqnarray}

\noindent
it assumes the form
\begin{eqnarray}
\sqrt{-y(1-y)}
\left ( {d \over dy} - {m/2 + B \over y} - {m/2 \over 1-y } \right ) R_{2} + \lambda R_{1} =0 \; ,
\nonumber
\\
\sqrt{-y(1-y)}
\left ( {d \over dy} + {m/2 + B \over y} + {m/2 \over 1-y } \right ) R_{1}
- \lambda R_{2} =0 \; .
\label{A.21}
\end{eqnarray}

First, let us  consider the cases   1 and $4'$:
\begin{eqnarray}
R_{1} = r_{1}\,y ^{-B - m/2} \; (1 - y)^{m/2} \;  F(a, b, c ;\; y ) \,, \nonumber
\\
a = -B +\sqrt{B^{2}-\lambda^{2}}\,,
\;
  b = -B
-\sqrt{B^{2}-\lambda^{2}} \, ,
\;
 c = -2B - m + {1\over 2}\, ;
\nonumber
\end{eqnarray}
\begin{eqnarray}
R_{2} = r_{2}\,y ^{-B - m/2+1/2} \; (1 - y)^{m/2+1/2 } \;  F(\alpha, \beta, \gamma ;\; y ) \, ,
\nonumber
\\
\alpha  = -B +1 +\sqrt{B^{2}-\lambda^{2}}=a+1\,, \nonumber
\\
  \beta = -B+1  -\sqrt{B^{2}-\lambda^{2}}=b+1 \, ,
  \nonumber
  \\ \gamma = -2B - m + {3\over 2}=c+1\,.
\nonumber
\end{eqnarray}

\noindent Substituting  expressions for $R_{1}$ and  $R_{2}$ into the second equation
\begin{eqnarray}
 \left (
{d \over dy} + {m/2 + B \over y} + {m/2 \over 1-y } \right ) R_{1} - { \lambda  \over \sqrt{-y(1-y)}} R_{2} =0 \; ,
\nonumber
\end{eqnarray}

\noindent that is
\begin{eqnarray}
 \left ( {d \over dy} + {m/2 + B \over y} + {m/2 \over 1-y } \right )
 r_{1} y ^{-B - m/2} \; (1 - y)^{m/2} \;  F(a, b, c ;\; y )
\nonumber
\\
-  i \lambda   r_{2}\,y ^{-B - m/2} \; (1 - y)^{m/2 } \;  F(a+1, b+1, c+1 ;\; y )
 =0 \; ,
\nonumber
\end{eqnarray}

\noindent we arrive at
\begin{eqnarray}
r_{1}  {d \over dy}  F(a,b,c,y)-
 i \lambda r_{2} F (a+1,b+1,c+1,y) = 0 \; ,
\nonumber
\end{eqnarray}

\noindent which means
\begin{eqnarray}
r_{1} {ab \over c} - i \lambda r_{2}=0 \; .
\label{A.22}
\end{eqnarray}

Now consider the cases  2 and  $3'$:
\begin{eqnarray}
R_{1} = r_{1}\,y ^{-B - m/2} \; (1 - y)^{1/2-m/2} \;  F(a', b', c' ;\; y ) \,, \nonumber
\\
a' = -B -m+{1\over 2}+\sqrt{B^{2}-\lambda^{2}}\,,
\nonumber
\\
  b' = -B-m+{1\over 2}
-\sqrt{B^{2}-\lambda^{2}} \, ,
\nonumber \\ c' = -2B - m + {1\over 2}\, ;
\nonumber
\end{eqnarray}
\begin{eqnarray}
R_{2} = r_{2}\,y ^{-B - m/2+1/2} \; (1 - y)^{-m/2 } \;  F(\alpha', \beta', \gamma' ;\; y ) \, ,
\nonumber
\\
\alpha'  = -B -m+{1\over 2} +\sqrt{B^{2}-\lambda^{2}}=a'\,,
\nonumber
\\
 \beta' = -B-m+{1\over 2} -\sqrt{B^{2}-\lambda^{2}}=b' \, ,
 \nonumber\\ \gamma' = -2B - m + {3\over 2}=c' +1 \,.
\nonumber
\end{eqnarray}

\noindent Substituting  expressions for $R_{1}$ and  $R_{2}$ into the second equation
\begin{eqnarray}
\left ( {d \over dy} + {m/2 + B \over y} + {m/2 \over 1-y } \right )
r_{1}y ^{-B - m/2}  (1 - y)^{1/2-m/2}   F(a', b', c' +1;\; y )
\nonumber
\\
- \lambda  r_{2}\, {y ^{-B - m/2+1/2} \; (1 - y)^{-m/2 }\over  \sqrt{-y(1-y)} }   F(a', b', c' ; y )  =0 \; .
\nonumber
\end{eqnarray}

\noindent we get
\begin{eqnarray}
r_{1} \left ( (m -1/2) + (1-y) {d \over d y} \right ) F(a', b', c' ; y ) -
i r_{2} F(a', b', c' +1; y ) = 0 \;.
\nonumber
\end{eqnarray}

\noindent
And further, with the use of identity
\begin{eqnarray}
- (a'+b'-c')  F(a',b',c',y)  +(1-y) {d \over dz} F(a',b',c',y)
\nonumber
\\
=
{(a'-c')(b'-c')\over c'} F(a',b',c'+1,y)
\label{formula}
\end{eqnarray}

\noindent
we arrive at  a needed relationship
\begin{eqnarray}
r_{1} {(a'-c')(b'-c')\over c'}  - i \lambda r_{2} F(a', b', c' +1; y ) =0\; .
\label{A.24}
\end{eqnarray}

In the next part of the paper, we consider an analogues  problem for a particle with spin $1/2$ described
 by Dirac equation in spherical Riemann  space. Though the treatment is similar, spherical
  geometry noticeably changes the task and final  results.

\section{ Cylindric coordinates and the Dirac equation \\ in spherical
 space  $S_{3}$, separation of the variables}

In the spherical Riemann  space $S_{3}$, let us use an extended
cylindric coordinates  (see \cite{Olevsky})
\begin{eqnarray}
dS^{2} =  dt^{2} -   \cos^{2} z ( d r^{2} + \sin^{2} r \; d
\phi^{2} ) + dz^{2}\; ]\; , \nonumber
\\
 z \in [-\pi /2 , + \pi /2
]\; , \qquad r \in [0, + \pi ] , \qquad \phi \in [0, 2 \pi ] \; ,
\nonumber
\\
u_{1} = \cos z  \sin r \cos \phi \; , \;  u_{2} = \cos z \; \sin r
\sin \phi \; ,
\nonumber
\\
 u_{3} = \sin z \; , \; u_{0} = \cos z  \cos r \;
. \label{2.4}
\end{eqnarray}

\noindent An analogue of usual homogeneous magnetic field is defined
as  \cite{4,5,6}
\begin{eqnarray}
 A_{\phi} = -2B \sin^{2} {r \over 2} = B\; ( \cos r -1 )\; .
\label{2.5}
\end{eqnarray}

\noindent
To coordinates  (\ref{2.4})  there corresponds the tetrad
\begin{eqnarray}
 e_{(a)}^{\beta}(x) = \left |
\begin{array}{llll}
1 & 0 & 0 & 0 \\
0 & \cos^{-1}z & 0 & 0 \\
0 & 0 & \cos^{-1}z\;\sin^{-1} r & 0 \\
0 & 0 & 0 & 1
\end{array} \right | \; .
\label{2.6}
\end{eqnarray}

 \noindent Christoffel symbols  $\Gamma^{r}_{\;\;jk }$ and Rici rotation coefficients
 $\gamma_{abc}$  are
\begin{eqnarray}
\Gamma^{r}_{\;\;jk } = \left | \begin{array}{ccc}
0 & 0 & -\mbox{tan}\;z \\
0 & - \sin r \cos r & 0 \\
- \mbox{tan}\;z & 0 & 0
\end{array} \right |  , \;
\Gamma^{\phi}_{\;\;jk } = \left | \begin{array}{ccc}
0 & \mbox{cot}\; r & 0\\
\mbox{cot}\; r & 0 &- \mbox{tan}\; z \\
0 & -\mbox{tan}\; z & 0
\end{array} \right |  ,
\nonumber
\end{eqnarray}
\begin{eqnarray}
 \Gamma^{z}_{\;\;jk } = \left | \begin{array}{ccc}
\sin z \cos z & 0 & 0\\
0 & \sin z \; \cos z \sin^{2} r & 0 \\
0 & 0 & 0
\end{array} \right | ,
\nonumber
\end{eqnarray}
\begin{eqnarray}
\gamma_{12 2} =
 { 1 \over \cos z \mbox{tan}\; r} \; , \;
 \gamma_{31 1} =
 -\mbox{tan}\; z\; , \; \gamma_{32 2} =
 -\mbox{tan}\; z\; .
 \label{2.8}
\end{eqnarray}

\noindent
A general covariant Dirac equation (for more detail see [10])
takes the form
\begin{eqnarray}
 [  \; i \gamma^{0} \partial_{t}  + {i \gamma^{1}  \over  \cos z}
 (    \partial_{r} + {1 \over 2}  {1 \over \mbox{tan}\;  r}  )
 + \gamma^{2}  {    i   \partial_{\phi} - e  B (\cos r -1)    \over \cos z \sin r}
 \nonumber
 \\
    +
 i \gamma^{3} ( \partial_{z} - \mbox{tan}\; z    ) - M  \;   ]  \Psi = 0\; .
\label{2.11}
\end{eqnarray}

\noindent
With the  substitution $ \Psi = \psi / \cos z  \sqrt{\sin r}
 $ eq. (\ref{2.11}) becomes simpler
\begin{eqnarray}
\left [  i \gamma^{1}  \partial _{ r }  +  \gamma^{2}
\;
 {i \partial_{\phi} - e B (\cos r -1)  \over \sin r }   +
  \cos z \left  (i  \gamma^{0} \partial _{ t}  +
  i  \gamma^{3}  \partial _{z} -  M  \right  )  \right ] \psi =0 \; .
\label{3.9}
\end{eqnarray}

Solutions of this equation will be searched in the form
\begin{eqnarray}
\psi = e^{-i\epsilon t} e^{im \phi} \left | \begin{array}{c}
f_{1}(r,z)\\ f_{2}(r,z)\\
f_{3}(r,z)\\
f_{4}(r,z) \
\end{array} \right | ;
\nonumber
\end{eqnarray}

\noindent so that
\begin{eqnarray}
\left [  i \gamma^{1}  \partial _{ r }  - \mu (r)
\gamma^{2}   +
  \cos z    ( \epsilon   \gamma^{0}   +
  i  \gamma^{3}  \partial_{ z} -  M    )  \right ]  \left | \begin{array}{c}
f_{1}(r,z)\\ f_{2}(r,z)\\
f_{3}(r,z)\\
f_{4}(r,z) \
\end{array} \right |
 =0 \; ,
 \label{3.10}
\end{eqnarray}

\noindent
 where
\begin{eqnarray}
 \mu (r) = { m -  e B ( 1 - \cos   r ) \over   \sin  r}
\;.
\nonumber
\end{eqnarray}
 Taking the  Dirac matrices in spinor basis, we get radial
equations for  $f_{a}(t,z)$
\begin{eqnarray}
  ( \partial_{ r }  + \; \mu  ) \; f_{4}
   + \cos z  \;   \partial_{z} \; f_{3}
+  i \; \cos z  \; ( \epsilon   f_{3}   -  M  f_{1} ) =0\,,
\nonumber
\\
( \partial _{ r }  - \; \mu  ) \;f_{3} -  \cos z  \;
 \partial _{z} \; f_{4}
 +   i  \; \cos z \; (  \epsilon  f_{4}   - M  f_{2} ) =0\,,
\nonumber
\\
 ( \partial _{ r }  +   \; \mu )  \; f_{2}
+ \cos z \;   \partial _{z} \; f_{1}
 - i   \; \cos z  \; ( \epsilon f_{1} -M f_{3} ) =0\,,
\nonumber
\\
  ( \partial _{r }  -  \; \mu  ) \; f_{1}
-  \cos z  \;  \partial _{z}\; f_{2} - i  \; \cos z
\; (   \epsilon f_{2} -M f_{4} ) =0\, . \label{3.15}
\end{eqnarray}

\noindent With linear restriction $ f_{3} = A f_{1} ,  \; f_{4} =
A f_{2}$, where
\begin{eqnarray}
\epsilon -{M \over A} = - \epsilon + M A \qquad  \Longrightarrow
\qquad A = A_{1,2}= {\epsilon \pm p \over M} \;   \label{3.18}
\end{eqnarray}

\noindent eqs. (\ref{3.15}) give
\begin{eqnarray}
 ( \partial_{ r }  +   \; \mu )  \; f_{2}
+ \cos z \;   \partial_{z}\;  f_{1}
 + i   \; \cos z  \; ( - \epsilon  + M  A  ) \;  f_{1} =0\,,
\nonumber
\\
  ( \partial _{ r }  -  \; \mu  ) \; f_{1}
-  \cos z  \;  \partial _{z}\; f_{2} + i  \; \cos z
\; (  -  \epsilon   + M  A ) \;  f_{2} =0\, . \label{3.19}
\end{eqnarray}

Thus, we have two possibilities

\vspace{4mm}
 $
 A =   (\epsilon + p)/M  \; ,
 $
\begin{eqnarray}
 ( \partial _{r }  +   \; \mu )  \; f_{2}
+ \cos z \;  (  \partial _{z} \;  +  i  p   \; )\;   f_{1} =0\,,
\nonumber
\\
  ( \partial _{r }  -  \; \mu  ) \; f_{1}
-  \cos z  \; (  \partial _{z}\;  - i  \; p\; )\;
f_{2} =0\, ; \label{3.20a}
\end{eqnarray}

$
 A =   (\epsilon - p)/M  \; ,
$
\begin{eqnarray}
 ( \partial _{ r }  +   \; \mu )  \; f_{2}
+ \cos z \;  (  \partial _{z}   -  i  p\;)   \;    f_{1} =0\,,
\nonumber
\\
  ( \partial _{r}   -  \; \mu  ) \; f_{1}
-  \cos z  \; (  \partial _{z}  + i  \; p\; ) \;
f_{2} =0\, . \label{3.20b}
\end{eqnarray}

Evidently, such a simplification can be rationalized through diaginalyzation  of extended
helicity operator $\Lambda$ (compare with (\ref{6.7}))
\begin{eqnarray}
\Lambda =   {1 \over \cos  z} \left (  \gamma^{2}\gamma^{3}\;  \partial _{r}    -i \;
\gamma^{3} \gamma^{1}  {i\partial _{\phi}  - e  B ( \cos r -1)  \over \sin  r } \right )
  +  \gamma^{1} \gamma^{2}  \partial _{z}    \; .
  \label{6.7'}
\end{eqnarray}

For definiteness, let us consider the system   (\ref{3.20a})
(transition to the case
 (\ref{3.20b}) is performed by the formal change $p \Longrightarrow -p$).
 Let us search solutions in the form
\begin{eqnarray}
f_{1} = Z_{1} (z) \; R_{1} (r) \; ,  \qquad  f_{2} = Z_{2}(z) \;
R_{2} (r)  \; . \nonumber
\end{eqnarray}

\noindent Introducing  a separating constant $\lambda$, we get two systems
\begin{eqnarray}
 \cos z  \; ({ d  \over d z} + i  \; p\; )\;
Z_{1}  = \lambda  \; Z_{2} \; ,
\qquad
 \cos z  \; ({ d  \over  d
z} - i  \; p\; )\; Z_{2} = \lambda \; Z_{1}  \; , \label{3.23}
\end{eqnarray}
\begin{eqnarray}
 ( {d \over d r }  +   \; \mu )  \;  R_{2} +
\lambda  \; R_{1}  =0\,,\qquad
   ( { d \over d r }  -  \; \mu  ) \;  R_{1}
-  \lambda   \; R_{2}  =0\, . \label{3.24}
\end{eqnarray}

\section{ Solution of the equations in  $z$-variable}

\vspace{2mm}

From  (\ref{3.23}) it follows the second-order differential
equations for $Z_{1}(z)$ and  $Z_{2}(z)$
\begin{eqnarray}
\left ( {d^{2} \over dz} - {\sin z \over \cos z} {d  \over dz}+
p^{2}-ip{\sin z\over \cos z}-{\lambda^{2}\over \cos^{2}z} \;
\right  )  Z_{1}=0\,,
\nonumber
\\
\left ( {d^{2} \over dz} - {\sin z \over \cos z} {d  \over dz}+
p^{2}+ip{\sin z\over \cos z}-{\lambda^{2}\over \cos^{2}z} \;
\right  )  Z_{2}=0\,.
 \label{3.25}
\end{eqnarray}

In a new variable  $
 y = (1 +  i\; \mbox{tan}\; z ) /  2$, with  the  use of the substitution $ Z_{1} = y^{A} (1-y)^{C} Z (y)$,
  eq. (\ref{3.25}) gives
\begin{eqnarray}
4y\,(1-y)\,{d^{2}Z\over dz^{2}}+4 [ \; 2A\,+\,{1\over
2}-(2A\,+\,2C\,+1)\,y \; ]{dZ\over dz} \nonumber
\\
+\;  [ \; {2A\,(2A-1)-p\,(p+1)\over y}+{2C\,(2C-1)-p\,(p-1)\over
1-y}\nonumber
\\
-4\,(A+C)^{2}  +4\lambda^{2} \; ] \; Z=0\,. \nonumber
\end{eqnarray}

\noindent With restrictions
\begin{eqnarray}
 A= -{p\over
2}\,,\; {p+1\over 2} \, , \qquad C = {p\over 2} \,,\;{1-p\over 2}
\, , \label{3.29}
\end{eqnarray}

\noindent for  $Z$  we get an equation of hypergeometric type \cite{Bateman}
\begin{eqnarray}
y (1-y){d^{2}Z\over dz^{2}}+\left[2A +{1\over
2}-(2A + 2C +1) y\right]{dZ\over
dz}-\left[(A+C)^{2}-\lambda^{2}\right]Z=0 \, , \nonumber
\end{eqnarray}

\noindent where
\begin{eqnarray}
a = A+C + \lambda \;, \qquad b = A+C  -\lambda \;, \qquad
c = 2A\,+\,{1\over 2} \; , \nonumber
\\
Z_{1} =  y^{A} \;( 1 - y )^{C} F(a,   b,
c ,    y  ) \; . \label{3.32}
\end{eqnarray}

Because $p >0$, to get solutions that are  finite  at the point $z = \pm \pi/2$ (that correspond
to the points $u_{a}=(0;0,0,\pm1)$ in the spherical space $S_{3}$), we should take negative
values for $A$ and $C$:
\begin{eqnarray}
A = -{p \over 2} , \qquad  C ={1- p \over 2}\; ,
\nonumber
\\
a =  -p + {1\over 2}  + \lambda \;, \; b = -p+{1 \over 2} -\lambda \;, \;
c = -p + {1\over 2} \; ,
\nonumber
\\
Z_{1}  = y^{A} (1-y)^{C} U_{1}(y)
= y^{-p/2}    (
1 - y )^{-p/2 +1/2} F(a,   b,
c ; \;  y )\; .
\label{3.33}
\end{eqnarray}

\noindent
Quantization condition is
\begin{eqnarray}
a= -n \qquad \Longrightarrow \qquad p = \lambda + n + {1 \over 2}\; .
\label{3.34}
\end{eqnarray}

\noindent Besides we must require
that
\begin{eqnarray}
A+B -n < 0 \qquad \Longrightarrow \qquad -p+n + {1\over 2}  < 0 \qquad \Longrightarrow \qquad \lambda >0 \;.
\nonumber
\label{3.33}
\end{eqnarray}

\noindent
There exist symmetrical quantization condition
\begin{eqnarray}
b= -n \qquad \Longrightarrow \qquad p = -\lambda + n + {1 \over 2} \; ,
\label{3.34}
\end{eqnarray}

\noindent and
\begin{eqnarray}
A+B -n < 0 \qquad \Longrightarrow \qquad -p+n + {1\over 2}  < 0
\qquad \Longrightarrow \qquad \lambda <0 \;.
\nonumber
\label{3.33}
\end{eqnarray}

\noindent
 These two possibilities are equivalent,
 for definiteness  we will take the variant with $\lambda > 0$.

Let us turn to the  equation for second function $Z_{2}$
\begin{eqnarray}
\left ( {d^{2} \over dz} -  {\sin z \over \cos z} {d  \over dz}+
p^{2} + ip{\sin z\over \cos z}-{\lambda^{2}\over \cos^{2}z} \;
\right  )  Z_{2}=0\,. \label{3.25'}
\end{eqnarray}

In a new variable  $
 y = (1 +  i\; \mbox{tan}\; z ) /  2$, with  the  use of the substitution $ Z_{2} = y^{K} (1-y)^{L} \bar{Z} (y)$,
  eq. (\ref{3.25'}) gives
\begin{eqnarray}
4y\,(1-y)\,{d^{2} \bar{Z}\over dz^{2}}+4 [ \; 2K\,+\,{1\over
2}-(2K\,+\,2L\,+1)\,y \; ]{d\bar{Z}\over dz} \nonumber
\\
+\;  [ \; {2K\,(2K-1) +p\,(-p+1)\over y}+{2L\,(2L-1) +p\,(-p-1)\over
1-y}\nonumber
\\
-4\,(K+L)^{2}  +4\lambda^{2} \; ] \; \bar{Z}=0\,. \nonumber
\end{eqnarray}

\noindent Requiring
\begin{eqnarray}
 K= +{p\over
2}\,,\; {-p+1\over 2} \, , \qquad L = -{p\over 2} \,,\;{1+p\over 2}
\, , \label{3.29'}
\end{eqnarray}

\noindent for  $\bar{Z}$  we get an equation of hypergeometric type
\begin{eqnarray}
y (1-y){d^{2} \bar{Z}\over d \bar{z}^{2}}+\left[2K +{1\over
2}-(2K + 2L +1) y\right]{d\bar{Z}\over
dz}-\left[(K+L)^{2}-\lambda^{2}\right] \bar{Z}=0 \, , \nonumber
\end{eqnarray}

\noindent where
\begin{eqnarray}
\alpha  = K+L + \lambda \;, \; \beta  = K+L  -\lambda \;, \;
\gamma  = 2K\,+\,{1\over 2} \; , \nonumber
\\
Z_{2} =  y^{K}   (
1 - y )^{L} F(\alpha,   \beta
\gamma ;     y ) \; . \label{3.32'}
\end{eqnarray}

Because $p >0$, to get solutions that are  finite  at the point $z = \pm \pi/2$, we should take negative
values for $K$ and $L$:
\begin{eqnarray}
K = {-p+1 \over 2} , \qquad  L ={- p \over 2}\; ,
\nonumber
\\
\alpha  =  -p + {1\over 2}  + \lambda \;, \; \beta  = -p+{1 \over 2} -\lambda \;, \;
\gamma  = -p + {3\over 2} \; ,
\nonumber
\\
Z_{2}  =
 y^{-p/2+1/2}   ( 1 - y )^{-p/2 } F(\alpha,  \; \beta,  \;
\gamma ;  y ) \; .
\nonumber
\end{eqnarray}

\noindent
Quantization condition is the old one
$$
\alpha = -n \qquad \Longrightarrow \qquad p = \lambda + n + {1 \over 2} \; .
$$

Now, it is the point to find a relative factor between
$Z_{1}(y)$ and $Z_{2}(y)$ with the help of the main first order system
\begin{eqnarray}
 i \left  (   y(1-y) { d  \over d y} +    {p \over 2} \right  )
Z_{1}  = \lambda  \sqrt{y(1-y) } \; Z_{2} \; ,
 \nonumber
 \\
 i \left  (  y(1-y) { d  \over d y}
 -   { p \over 2} \right  ) Z_{2} = \lambda   \sqrt{y(1-y)} \; Z_{1}  \; . \label{3.23''}
\end{eqnarray}

\noindent
Starting  with
\begin{eqnarray}
A=-{p\over 2}\,,\qquad C={1\over 2}-{p\over 2}\,,
\nonumber
\\
Z_{1}=z_{1}\,y^{-p/2}\,(1-y)^{1/2-p/2}\,F\,(a,\,b,\,c;\;y)\,,
\nonumber
\\
a=-p+{1\over 2}+\lambda\,,\qquad b=-p+{1\over 2}-\lambda\,,\qquad c=-p+{1\over 2}\,;
\nonumber
\\
K=-{p\over 2}+{1\over 2}\,,\qquad L=-{p\over 2}\,,
\nonumber
\\
Z_{2}=z_{2}\,y^{-p/2+1/2}\,(1-y)^{-p/2}\,F\,(\alpha,\,\beta,\,\gamma;\;y)\,,
\nonumber
\\
\alpha=-p+{1\over 2}+\lambda=a\,,\;  \beta=-p+{1\over 2}-\lambda=b\,,\; \gamma=-p+{3\over 2}=c+1\,,
\nonumber
\end{eqnarray}

\noindent and substituting functions  $Z_{1}$ and $Z_{2}$ into the first equation in (\ref{3.23''}),
we get
\begin{eqnarray}
z_{1} \left [ (1-y)\,{d \over  d y} - (a+b-c)   \right ] \; F(a,\,b,\,c;\;y)\,z_{1}
 +i \lambda\, \,z_{2} F(a,\,b,\,c+1;\;y)=0\, .
\nonumber
\end{eqnarray}

\noindent With the use of the known relationship
\begin{eqnarray}
- (a+b-c)  F(a,b,c,y)  +(1-y) {d \over dz} F(a,b,c,y)
=
{(a-c)(b-c)\over c} F(a,b,c+1,y) \; ,
\label{formula}
\end{eqnarray}

\noindent it gives
\begin{eqnarray}
z_{1} {(a-c)(b-c)\over c} F(a,b,c+1,y) +
 +i \lambda\, \,z_{2} F(a,\,b,\,c+1;\;y)=0\,,
\nonumber
\end{eqnarray}

\noindent that is
\begin{eqnarray}
z_{1} {(a-c)(b-c)\over c}   + i \lambda\, \,z_{2} =0\,.
\end{eqnarray}

\section{ Solution of the equations in  $r$-variable}

From eqs. (\ref{3.24}) it follows a second-order differential
equation for $R_{1}$ (for brevity let  $eB$  be  noted as $B$)
\begin{eqnarray}
{d^{2}R_{1}\over dr^{2}}+\left[{m\,\cos  r-B\,(\cos  r-1)\over
\sin^{2}  r }-{[m+B\,(\cos  r-1)]^{2}\over \sin^{2}  r
}+\lambda^{2}\right]R_{1}=0\,.
\label{3.35}
\end{eqnarray}

\noindent With  a new variable $ y= (1+ \cos r )/ 2 $,  eq.
(\ref{3.35}) reads
\begin{eqnarray}
y(1-y){d^{2}R_{1}\over dy^{2}}+ ({1\over
2}-y ){dR_{1}\over dy}
- \left [ \; -\lambda^{2}+{m^{2}\over 4} ({1\over y}+ {1\over
1-y} ) \right.
\nonumber
\\
\left. +{m\over 4}  ({1\over y}-{1\over 1-y} )
 -{mB
\over y}-B^{2} (1-{1\over y} )-{B\over
2y} \right ]  R_{1}=0 \; .
\nonumber
\label{3.35'}
\end{eqnarray}

\noindent With the substitution $ R_{1} = y^{A} (1-y)^{C} \bar{R}_{1} (y)$,
we get
\begin{eqnarray}
y(1-y){d^{2} \bar{R}_{1} \over dy^{2}}+ \left [  2A + {1\over 2}-(2A+2C +1)\;
y  \right] \,{d \bar{R}_{1}\over dy}  \nonumber
\\
+\left [ { A^{2}-A/2-m^{2}/4-m/4+ mB - B^{2}+B/2\over y} \right.
\nonumber
\\
\left.+{C^{2}-C/2- m^{2}/4 + m/4\over 1-y}
-(A+C)^{2}+\lambda^{2}+B^{2}\right] \bar{R}_{1}=0\, . \nonumber
\end{eqnarray}

\noindent Requiring
\begin{eqnarray}
 A= {2B - m\over 2}\; ,\;  {1 -(2B - m)  \over 2} \, , \qquad
 C = {m\over 2} ,\; {1 -m \over 2} \, ,
 \label{3.37}
 \end{eqnarray}

\noindent we arrive at a differential equation of hypergeometric
type \cite{Bateman}
\begin{eqnarray}
y(1-y){d^{2} \bar{R}_{1}\over dy^{2}}+   [  2A + {1\over 2}-(2A+2C +1)
y  ] {d\bar{R}_{1}\over dy}
\nonumber
\\
- [ \; (A+C)^{2}-\lambda^{2}-B^{2} \;  ] \; \bar{R}_{1}=0\, . \label{3.38}
\end{eqnarray}

\noindent where
\begin{eqnarray}
\alpha = A+C - \sqrt{B^{2}+\lambda^{2}}\;, \; \beta = A+C
+\sqrt{B^{2}+\lambda^{2}} \;,  \nonumber
\\
\gamma = 2A+ 1 /  2\;, \qquad
R_{1} = y ^{A} \; (1 - y)^{C} \; F(\alpha,
\beta, \gamma ;\; y ) \; .
 \label{3.39}
 \end{eqnarray}

In order to have a finite  solution at the origin $r=0$ (that
corresponds to  the half-curve
 $u_{0} = + \cos z,
\;  u_{3} = \sin  z  ,  u_{1}=0 ,  u_{2}=0$) and at  $r = \pi$
(that corresponds to the other part of the curve,  $ u_{0} = -
\cos z  ,   u_{3} = \sin  z  ,  u_{1}=0 , u_{2}=0$ ),  we  must
take positive values for $A$ and  $C$ ($  A > 0 \; ,
\;  \; C > 0 $)
\begin{eqnarray}
R_{1} = y ^{A}  (1 - y)^{C}   F(\alpha,
\beta, \gamma ; \; y) \;   . \label{3.41}
\end{eqnarray}

\noindent
There are possible  four variants  (for definiteness
assuming   $B>0$):
\begin{eqnarray}
\mbox{variant 1 } \; ,\qquad   A
=
 {2B - m  \over 2}  \;, \qquad C = {1-m \over 2}  \;   ,
\nonumber
\\
\mbox{variant } \; 2\; , \qquad
 A =
 {2B - m  \over 2}    \; , \qquad C = {m \over 2}  \; ,
\nonumber
\\
\mbox{variant } \; 3 \; , \qquad  A =
{ m +1 - 2B \over 2}  \; , \qquad  C = {m \over 2}   \; ,
\nonumber
\\[5mm]
\mbox{variant } \; 4\; ,  \qquad   A
= { m +1 - 2B \over 2}     \;  , \qquad  C = {1-m \over 2}   \; .
\nonumber
\label{3.42}
\end{eqnarray}

For definiteness, let us consider $B>0$.
There arise the  following three types of solutions

\vspace{2mm}

\underline{$ \mbox{Variant } \; 1\; , \qquad   m  = ..., -3/2 , -1/2, +1/2   \; $,}
\begin{eqnarray}
R_{1} = y ^{(2B-m)/2}  \; (1 -  y)^{(1-m)/2} \;
F(a, b, c ;\; y ) \; ,
\nonumber
\\
a =  B - m +1/2  - \sqrt{\lambda^{2} + B^{2}} = -n  \;,
\nonumber
\\
 b =   B-m +  1/2 + \sqrt{\lambda^{2} + B^{2}} \; ,\;\;
 c = 2B  - m + {1\over 2}
\; , \nonumber
\\
\mbox{spectrum} \qquad \sqrt{\lambda^{2} +B^{2} } = n - m +1/2 + B   \; .
 \label{3.43}
 \end{eqnarray}

\underline{$ \mbox{Variant  } \; 2\; , \qquad  m = +1/2, ...,  \mu  < 2B ,$}
\begin{eqnarray}
R_{1} = y ^{(2B-m)/2} \; (1 -  y)^{m/2} \;
F(a', b', c' ;\; y ) \; ,
\nonumber
\\
a' = B - \sqrt{\lambda^{2} + B^{2}}= -n  \;,
\nonumber
\\
  b'  =  B + \sqrt{\lambda^{2} + B^{2}}\;,
  \;\;
    c' =2B - m + 1/2
 \; ; \nonumber
\\
\mbox{spectrum} \qquad
 \sqrt{ \lambda^{2} +B^{2}}   =  B+n  \;
\label{3.44}
\end{eqnarray}

\vspace{2mm}

\underline{$ \mbox{Variant } \; 3\; , \qquad  m >
2B -1 \;$ } ,
\begin{eqnarray}
R_{1} = y ^{(m+1 - 2B)/2 } \; (1 - y)^{m/2} \;
F(a'', b'', c'' ;\;y ) \; ,
\nonumber
\\
a'' = m+1/2 -B  - \sqrt{\lambda^{2} + B^{2}} = - n'' \;,
\nonumber
\\
b'' =  m+1/2 - B
 + \sqrt{\lambda^{2} + B^{2}}\;,
 \;\;
  c'' = - 2B + m +3/2 \; .
\nonumber
\\
\mbox{spectrum}\qquad
\sqrt{\lambda^{2} +B^{2}} = n'' +m +1/2 - B \; . \label{3.45}
\end{eqnarray}

From eqs. (\ref{3.24}) it follows a second-order differential
equation for $R_{2}$ (again let  $eB$  be  noted as $B$)
\begin{eqnarray}
\left ( {d^{2}R_{2}\over dr^{2}} - {m\,\cos  r-B\,(\cos  r-1)\over
\sin^{2}  r }-{[m+B\,(\cos  r-1)]^{2}\over \sin^{2}  r
}+\lambda^{2}\right) R_{2}=0\,.
\label{3.35'}
\end{eqnarray}

\noindent With  a new variable $ y= (1+ \cos r )/ 2 $,  and the substitution $ R_{2} = y^{A} (1-y)^{C} \bar{R}_{2} (y)$,
we get
\begin{eqnarray}
y(1-y){d^{2} \bar{R}_{2}\over dy^{2}}+ \left [  2K + {1\over 2}-(2K+2L +1)\;
y  \right] \,{d \bar{R}_{2} \over dy}  \nonumber
\\
+\left [ { K^{2}- K/2-m^{2}/4 +m/4 + m B - B^{2} -B/2\over y} \right.
\nonumber
\\
\left.+{K^{2}-L/2- m^{2}/4 - m/4\over 1-y}
-(K+L)^{2}+\lambda^{2}+B^{2}\right] \bar{R}_{2}=0\, . \nonumber
\end{eqnarray}

\noindent With restrictions
\begin{eqnarray}
 K= - {2B - m\over 2}\; ,\;  {1 +(2B - m)  \over 2} \, , \qquad
 L = -{m\over 2} ,\; {1 +m \over 2} \, ,
 \label{3.37'}
 \end{eqnarray}

\noindent we arrive at a differential equation of hypergeometric
type \cite{Bateman}
\begin{eqnarray}
y(1-y){d^{2} \bar{R}_{2} \over dy^{2}}+   [  2K + {1\over 2}-(2K+2L +1)
y  ] {d\bar{R}_{2}\over dy}
\nonumber
\\
- [ \; (K+L)^{2}-\lambda^{2}-B^{2} \;  ] \; \bar{R}_{2}=0\, , \label{3.38'}
\end{eqnarray}

\noindent where $\gamma = 2K+ 1 /  2$ and
\begin{eqnarray}
\alpha = K+L - \sqrt{B^{2}+\lambda^{2}}\;, \; \beta = K+L
+\sqrt{B^{2}+\lambda^{2}} \;,  \nonumber
\\
R_{2} = y ^{K} \; (1 - y)^{L} \; F(\alpha,
\beta, \gamma ;\; y ) \; .
 \label{3.39'}
 \end{eqnarray}

To have a finite  solution at the origin $r=0$ (that
corresponds to  the half-curve
 $u_{0} = + \cos z,
\;  u_{3} = \sin  z  ,  u_{1}=0 ,  u_{2}=0$) and at  $r = \pi$
(that corresponds to the other part of the curve,  $ u_{0} = -
\cos z  ,   u_{3} = \sin  z  ,  u_{1}=0 , u_{2}=0$ ),  we  must
take positive values for $K$ and  $L$ ($  K > 0 \; ,
\;  \; L > 0 $)
\begin{eqnarray}
R_{1} = y ^{A}  (1 - y)^{C}   F(\alpha,
\beta, \gamma ; \; y) \;   . \label{3.41'}
\end{eqnarray}

\noindent
There are possible  four variants  (assuming   $B>0$):
\begin{eqnarray}
\mbox{variant} 1'  \; ,\qquad   K
=
 {-2B + m  \over 2}  \;, \qquad L = {1+m \over 2}  \;   ,
\nonumber
\\
\mbox{variant } \; 2'\; , \qquad
 K =
 {-2B + m  \over 2}    \; , \qquad L = {-m \over 2}  \; ,
\nonumber
\\
\mbox{variant } \; 3' \; , \qquad  K =
{ -m +1 + 2B \over 2}  \; , \qquad  L = {-m \over 2}   \; ,
\nonumber
\\
\mbox{variant } \; 4'\; ,  \qquad   K
= { -m +1 + 2B \over 2}     \;  , \qquad  L = {1+m \over 2}   \; .
\nonumber
\label{3.42'}
\end{eqnarray}

There are possible three of them.
\begin{eqnarray}
\underline{\mbox{Variant} 1'}\;,  \qquad \qquad m > 2B \;,
\nonumber
\\
R_{2} = y^{-B+m/2} (1-y)^{m/2+1/2} F(\alpha', \beta', \gamma',y) \; ,
\nonumber
\\
\alpha' = -B+m+1/2 - \sqrt{B^{2}+\lambda^{2}} \; ,
\nonumber
\\
\beta' = -B+m+1/2 + \sqrt{B^{2}+\lambda^{2}} \; ,
\nonumber
\\
\gamma' = -2B + m +1/2 \; ,
\end{eqnarray}

\noindent with quantization condition
\begin{eqnarray}
\alpha' = - n\;, \qquad  \sqrt{B^{2}+\lambda^{2}} = n +m+1/2 - B   > 0\; ,
\end{eqnarray}

\noindent
which correlates with the \underline{variant 3}  for $R_{1}(y)$.
\begin{eqnarray}
\underline{\mbox{Variant} 3'}\;,  \qquad \qquad m <0 \;,
\nonumber
\\
R_{2} = y^{B-m/2+1/2} (1-y)^{-m/2} F(\alpha'', \beta'', \gamma'',y) \; ,
\nonumber
\\
\alpha'' = B-m+1/2 - \sqrt{B^{2}+\lambda^{2}} \; ,
\nonumber
\\
\beta'' = B-m+1/2 + \sqrt{B^{2}+\lambda^{2}} \; ,
\nonumber
\\
\gamma'' = 2B - m +3/2 \; ,
\end{eqnarray}

\noindent with quantization condition
\begin{eqnarray}
\alpha'' = - n\;, \qquad  \sqrt{B^{2}+\lambda^{2}} = n -m+1/2 + B   > 0\;
\end{eqnarray}

\noindent
which correlates with the \underline{variant 1}  for $R_{1}(y)$.
\begin{eqnarray}
\underline{\mbox{Variant}\;\;  4'}\;,  \qquad \qquad -1 < m < 1+2B  \;,
\nonumber
\\
R_{2} = y^{B-m/2+1/2} (1-y)^{m/2+1/2 } F(\alpha''', \beta''', \gamma''',y) \; ,
\nonumber
\\
\alpha''' = B +1  - \sqrt{B^{2}+\lambda^{2}} \; ,
\nonumber
\\
\beta''' = B +1  + \sqrt{B^{2}+\lambda^{2}} \; ,
\nonumber
\\
\gamma''' = 2B - m +3/2 \; ,
\end{eqnarray}

\noindent with quantization condition
\begin{eqnarray}
\alpha''' = - n\;, \qquad  \sqrt{B^{2}+\lambda^{2}} = n +1 + B   \;,
\end{eqnarray}

\noindent
which correlates with the \underline{variant 2}  for $R_{1}(y)$.

Now we should find relative  factor between $R_{1}$ and $R_{2}$.
In the variable $y = (1 + \cos r)/2$, the relevant  radial system
 assumes the form
\begin{eqnarray}
-\sqrt{y(1-y)} \left ( {d \over  dy} + {B-m/2 \over y} - {m/2 \over  1-y} \right )R_{2}  + \lambda R_{1}=0\;,
\nonumber
\\
-\sqrt{y(1-y)} \left ( {d \over  dy} - {B-m/2 \over y} + {m/2 \over  1-y} \right )R_{1}  - \lambda R_{2}=0\; .
\end{eqnarray}

Let us find relative factor for $R_{1}, R_{2}$ in the variants $1--\;3'$:

\vspace{3mm}

\underline{$ \mbox{Variant } \; 1\; , \qquad   m < 0\; $,}
\begin{eqnarray}
R_{1} = y ^{(2B-m)/2}  \; (1 -  y)^{(1-m)/2} \;
F(a, b, c ;\; y ) \; ,
\nonumber
\\
a =  B - m +1/2  - \sqrt{\lambda^{2} + B^{2}}   \;,
\nonumber
\\
 b =   B-m +  1/2 + \sqrt{\lambda^{2} + B^{2}} \; ,
 \nonumber
 \\
 c = 2B  - m + {1\over 2}\; ;
 \nonumber
 \end{eqnarray}

$\underline{\mbox{Variant} 3'}\;,  \qquad \qquad m <0 \;,
$
\begin{eqnarray}
R_{2} = y^{B-m/2+1/2} (1-y)^{-m/2} F(\alpha'', \beta'', \gamma'',y) \; ,
\nonumber
\\
\alpha'' = B-m+1/2 - \sqrt{B^{2}+\lambda^{2}} = a \; ,
\nonumber
\\
\beta'' = B-m+1/2 + \sqrt{B^{2}+\lambda^{2}}  = b \; ,
\nonumber
\\
\gamma'' = 2B - m +3/2  = c +1 \; .
\nonumber
\end{eqnarray}

Substituting these functions into the second equation
\begin{eqnarray}
-\sqrt{y(1-y)} \left ( {d \over  dy} - {B-m/2 \over y} + {m/2 \over  1-y} \right )
\nonumber
\\
\times
r_{1} y ^{(2B-m)/2}  (1 -  y)^{(1-m)/2} F(a, b, c ; y )
\nonumber
\\
 - \lambda  r_{2}
y^{B-m/2+1/2} (1-y)^{-m/2} F(a, b, c+1,y) =0\; .
\nonumber
\end{eqnarray}

\noindent we get
\begin{eqnarray}
r_{1} \left ( (B-{m\over 2}){1\over y}(1-y) - ({1\over 2} - {m\over 2}) +(1-y){d \over dy}\right.
\nonumber
\\
\left.  -
(B-{m\over 2}){1\over y}(1-y) + {m \over 2}  \right )F(a, b, c ; y )
+
\lambda r_{2} F(a, b, c+1,y) =0 \; .
\nonumber
\end{eqnarray}

\noindent
It reads differently as
\begin{eqnarray}
r_{1} [\; (c-a-b) +(1-y) {d \over dy} \; ]\;  F(a, b, c ; y ) + \lambda r_{2} F(a, b, c+1,y) =0 \; ;
\nonumber
\end{eqnarray}

\noindent from whence with the use of the formula
\begin{eqnarray}
(c-a-b)  F(a,b,c,y)  +(1-y) {d \over dz} F(a,b,c,y)
=
{(a-c)(b-c)\over c} F(a,b,c+1,y) \; ,
\nonumber
\end{eqnarray}

\noindent we arrive at a needed relationship
\begin{eqnarray}
{(a-c)(b-c)\over c}  r_{1} + \lambda r_{2} =0\;.
\end{eqnarray}

Let us find relative factor for $R_{1}, R_{2}$ in the variants $2--\;4'$:

\vspace{3mm}

\underline{$ \mbox{Variant  } \; 2\; , \qquad  m = +1/2, ...,  \mu  < 2B ,$}
\begin{eqnarray}
R_{1} = y ^{(2B-m)/2} \; (1 -  y)^{m/2} \;F(a', b', c' ;\; y ) \; ,
\nonumber
\\
a' = B - \sqrt{\lambda^{2} + B^{2}} \;,
\nonumber
\\
  b'  =  B + \sqrt{\lambda^{2} + B^{2}}\;,
  \nonumber
  \\
    c' =2B - m + 1/2\; ;
 \; ; \nonumber
\end{eqnarray}
\begin{eqnarray}
\underline{\mbox{Variant}\;\;  4'}\;,  \qquad \qquad -1 < m < 1+2B  \;,
\nonumber
\\
R_{2} = y^{B-m/2+1/2} (1-y)^{m/2+1/2 } F(\alpha''', \beta''', \gamma''',y) \; ,
\nonumber
\\
\alpha''' = B +1  - \sqrt{B^{2}+\lambda^{2}} =a' +1  \; ,
\nonumber
\\
\beta''' = B +1  + \sqrt{B^{2}+\lambda^{2}}  =b' +1 \; ,
\nonumber
\\
\gamma''' = 2B - m +3/2  =c' +1 \; .
\nonumber
\end{eqnarray}

Substituting these functions into the second equation
\begin{eqnarray}
-\sqrt{y(1-y)} \left ( {d \over  dy} - {B-m/2 \over y} + {m/2 \over  1-y} \right )
\nonumber
\\
\times
r_{1} y ^{(2B-m)/2} \; (1 -  y)^{m/2} \;F(a', b', c' ;\; y )
\nonumber
\\
  - \lambda r_{2} y^{B-m/2+1/2} (1-y)^{m/2+1/2 } \;F(a'+1, b'+1, c' +1;\; y )=0\; .
\nonumber
\end{eqnarray}

\noindent we get
\begin{eqnarray}
r_{1} \left ( {B-m/2 \over y} - {m/2 \over 1-y} +{d \over dy} -
{B-m/2 \over y}+ {m/2 \over 1-y} \right ) F(a', b', c' ;\; y )
\nonumber
\\
+ \lambda r_{2} y^{B-m/2+1/2} (1-y)^{m/2+1/2 } \;F(a'+1, b'+1, c' +1;\; y )=0\; ;
\nonumber
\end{eqnarray}

\noindent from whence it follows
\begin{eqnarray}
r_{1} {a' b' \over c'} + r_{2} \lambda =0 \; .
\end{eqnarray}

Let us find a relative factor for $R_{1}, R_{2}$ in the variants $ 3--\;4'$:

\underline{$ \mbox{Variant } \; 3\; , \qquad  m >
2B -1 \;$ } ,
\begin{eqnarray}
R_{1} = y ^{(m+1 - 2B)/2 } \; (1 - y)^{m/2} F(a'', b'', c'' ;\;y ) \; ,
\nonumber
\\
a'' = m+1/2 -B  - \sqrt{\lambda^{2} + B^{2}} = \alpha '   \;,
\nonumber
\\
b'' =  m+1/2 - B
 + \sqrt{\lambda^{2} + B^{2}} =\beta ' \;,
 \nonumber
 \\
  c'' = - 2B + m +3/2  = \gamma '+1 \; .
\nonumber
\end{eqnarray}
\begin{eqnarray}
\underline{\mbox{Variant} 1'}\;,  \qquad \qquad m > 2B \;,
\nonumber
\\
R_{2} = y^{-B+m/2} (1-y)^{m/2+1/2} F(\alpha', \beta', \gamma',y) \; ,
\nonumber
\\
\alpha' = -B+m+1/2 - \sqrt{B^{2}+\lambda^{2}} \; ,
\nonumber
\\
\beta' = -B+m+1/2 + \sqrt{B^{2}+\lambda^{2}} \; ,
\nonumber
\\
\gamma' = -2B + m +1/2 \; .
\end{eqnarray}

Substituting these functions into the first equation
\begin{eqnarray}
-\sqrt{y(1-y)} \left ( {d \over  dy} + {B-m/2 \over y} - {m/2 \over  1-y} \right )
\nonumber
\\
\times \;
r_{2} \; y^{-B+m/2} (1-y)^{m/2+1/2} F(\alpha', \beta', \gamma',y)
\nonumber
\\
 + \lambda r_{1} y ^{(m+1 - 2B)/2 } \; (1 - y)^{m/2} F(\alpha', \beta', \gamma' +1 ;\;y )=0\;,
\nonumber
\end{eqnarray}

\noindent we get
\begin{eqnarray}
-r_{2} \left ( (-B +m/2 ){1-y \over y} -(m/2 +1/2)+ (1-y){d \over dy}
\right.
\nonumber
\\
\left.
+(B-m/2){1-y \over y} -m/2 \right ) F(\alpha', \beta', \gamma',y)
+ \lambda r_{1}  F(\alpha', \beta', \gamma' +1 ;\;y )=0\;,
\nonumber
\end{eqnarray}

\noindent or differently
\begin{eqnarray}
- r_{2} \left ( ( \gamma' - \alpha' - \beta') +(1-y){d \over d y}\right ) F(\alpha', \beta', \gamma',y)
\nonumber
\\
+ \lambda r_{1}  F(\alpha', \beta', \gamma' +1 ;\;y )=0\;,
\nonumber
\end{eqnarray}

\noindent which results in
\begin{eqnarray}
- r_{2} {(\alpha' - \gamma')(\beta' - \gamma') \over \gamma '} +  \lambda r_{1} =0
\end{eqnarray}

\section{Conclusions}

Let us summarize results.

First,  note that  in Lobachevsky space, the formulas  (\ref{A.40}) describe finite number of discrete energy levels,  governed
 by the magnitude of magnetic field $B$. The whole situation  can be clarified by Fig.1.

\vspace{-15mm}

\unitlength=0.56mm
\begin{picture}(160,100)(-120,0)

\special{em:linewidth 0.4pt} \linethickness{0.4pt}

\put(-70,0){\vector(+1,0){140}}  \put(+70,-5){$m$}
\put(0,-50){\vector(0,+1){100}}   \put(+2,+53){$\mid m \mid -\mid 2B +m \mid  +  2n <0$}

\put(0,0){\line(+1,+1){40}}
\put(0,0){\line(-1,+1){40}}
\put(+45,+35){$\mid m \mid $}

\put(-30,0){\circle*{2}}  \put(-35,+2){$-2B$}

\put(-15,0){\circle*{2}}   \put(-15,+2){$-B$}

 \put(0,+30){\circle*{2}}  \put(2,+30){$+2B$}

\put(-30,0){\line(+1,-1){70}}  \put(-30,0){\line(-1,-1){30}}  \put(-95,-20){$-\mid 2B +m \mid $}

\put(-15,+0.2){\line(+1,0){85}}
\put(-15,+0.1){\line(+1,0){85}}
\put(-15,-0.1){\line(+1,0){85}}
\put(-15,-0.2){\line(+1,0){85}}

\end{picture}

\vspace{45mm}

\begin{center}
{\bf Fig. 1.  $H_{3}$-model,  bound states at $B>0, \; -B < m $ }
\end{center}

\vspace{5mm}

Besides, results  from (\ref{A.40}) may be presented by an unifying relation
\begin{eqnarray}
A+C +\sqrt{B^{2}- \lambda^{2} } = - n \qquad \Longrightarrow
\nonumber
\\
\sqrt{B^{2}- \lambda^{2} } = - {\mid 2B + m \mid  \over  2} + {\mid m \mid \over 2} + n \; .
\label{main-1}
\end{eqnarray}

This form is valid for negative $B$  as well,  though Fig. 1 should be modified for $B<0$.

Note here the way of transition to the limit of  the flat Minkowski space -- it  is
realized as follows
\begin{eqnarray}
\lambda ^{2} \rightarrow {P ^{2}_{z} \rho^{2}  \over \hbar^{2}  }
= \lambda_{0}^{2} \; \rho^{2} \; ,\qquad B \rightarrow {e B \over
\hbar c}\;  \rho^{2} \; , \nonumber
\\
1. \qquad \lambda^{2}_{0}  =   {2e B \over \hbar } \; n   \; ;
\qquad 2. \qquad \lambda^{2}_{0}  =   {2e B \over \hbar }\;
(n -m +{1\over 2})   \; . \label{A.41}
\end{eqnarray}

In the spherical Riemann model, similarly results on quantization (\ref{3.43})--(\ref{2.45})  can be
 presented by an unifying formula
\begin{eqnarray}
A+C - \sqrt{B^{2}- \lambda^{2} } = - n \qquad \Longrightarrow
\nonumber
\\
\sqrt{B^{2}- \lambda^{2} } =  {\mid 2B - m \mid  \over  2} + {\mid m \mid \over 2} + n \; .
\label{main-1'}
\end{eqnarray}

\noindent
and  Fig. 2

\vspace{-15mm}

\unitlength=0.56mm
\begin{picture}(160,100)(-120,0)

\special{em:linewidth 0.4pt} \linethickness{0.4pt}

\put(-70,0){\vector(+1,0){140}}  \put(+70,-5){$m$}
\put(0,-50){\vector(0,+1){100}}   \put(+2,+53){$\mid m \mid -\mid 2B -m \mid  +  2n >0$}

\put(0,0){\line(+1,+1){40}}
\put(0,0){\line(-1,+1){40}}
\put(-55,+35){$\mid m \mid $}

\put(+30,-7){\circle*{2}}  \put(+35,+2){$+2B$}

 \put(0,+30){\circle*{2}}  \put(2,+30){$+2B$}

\put(+30,0){\line(+1,+1){40}}  \put(+30,0){\line(-1,+1){40}}  \put(+70,+25){$+\mid 2B -m \mid $}


\end{picture}

\vspace{30mm}

\begin{center}
{\bf Fig. 2.  $S_{3}$-model,  bound states at $B>0$ }
\end{center}

\vspace{5mm}

Evidently, in the case of negative $B$ this Fig.2 should be slightly modified.

\vspace{5mm}

There should be given a  clarifying remarks about quantum number $m$. In fact,
the above used  relationship $-i \partial_{\phi } \Psi = m  \;
\Psi $  (in both curved models) represents transformed from Cartesian coordinates to
cylindrical an eigenvalue  equation for the third projection of
the the total angular momentum of the Dirac particle
\begin{eqnarray}
\hat{J}_{3} \Psi_{Cart}  = ( -i {\partial \over  \partial \phi }
  + \Sigma_{3} )\; \Psi_{Cart} =  m\; \Psi_{Cart} \; ;
\label{3.45}
\end{eqnarray}

\noindent  this means that for  the quantum number
 $m$ are permitted only  half-integer values $ m = \pm {1 \over 2}, \; \pm {3\over 2}, ... $.

Note that some other new problems
concerning quantum mechanical particles in  external magnetic field< in flat Minkowski space and in curved spaces
of constant
curvature, were considered in \cite{Book-4}.
More wide discussion of the different quantum mechanical problems in spaces of Lobachevsky $H_{3}$ and Riemann $S_{3}$
may be found in \cite{Book-5}.

\section{Acknowledgement}

Authors are  grateful to participants of the Scientific Seminar of
the Laboratory of theoretical physics, Institute of physics of
National Academy od Sciences of Belarus, for discussion.

This  work was   supported   by the Fund for Basic Researches of Belarus,
 F11M-152.

\end{document}